\documentclass[prd,showpacs,nofootinbib,amsmath,amssymb,superscriptaddress,twocolumn]{revtex4-2}

\usepackage{graphicx}
\usepackage{dcolumn}
\usepackage{bm}
\usepackage{amsmath,amssymb,bbm,mathrsfs,nicefrac}
 \usepackage{array,multirow}
 \usepackage{float}
\usepackage{tikz}
\usepackage{amsmath}
\usepackage{amssymb}
\usepackage{graphicx,textcomp,float,gensymb,wrapfig, enumitem,comment,dsfont,framed,slashed,appendix,wrapfig,xcolor}
\usepackage{tabularray}
\UseTblrLibrary{booktabs}
\usepackage{ bbold }
\usepackage{braket} 
\usepackage{ mathrsfs }
\usepackage{etoolbox,hyperref,makecell}
\usepackage{cleveref}
\usepackage{orcidlink}

\hypersetup{
	colorlinks=true,
	citecolor=blue,
	citebordercolor=red,
	linktoc=all,
	linkcolor=blue,
	urlcolor=blue
}

\AddToHook{cmd/appendix/before}{
    
    \crefalias{section}{appendix}
    \crefalias{subsection}{appendix}
}

\crefformat{equation}{#2Eq.~(#1)#3}
\crefformat{appendix}{#2App.~#1#3}
\crefformat{section}{#2Sec.~#1#3}
\crefformat{table}{#2Tab.~#1#3}
\crefformat{figure}{#2Fig.~#1#3}

\crefmultiformat{section}{#2Secs.~#1#3}{ and~#2#1#3}{, #2#1#3}{ and~#2#1#3}
\crefmultiformat{appendix}{#2Apps.~#1#3}{ and~#2#1#3}{, #2#1#3}{ and~#2#1#3}

\crefrangeformat{equation}{#3Eqs.~(#1)#4 to #5(#2)#6}

\Crefformat{figure}{#2Fig.~#1#3}

\usepackage{feynmp}
\usetikzlibrary{arrows.meta}
\usepackage{empheq}
\usepackage[normalem]{ulem}
\usepackage{scrextend}
\usepackage{slashed}
\newcommand{\bea}{\begin{eqnarray}}  
\newcommand{\eea}{\end{eqnarray}}

\definecolor{darkgreen}{rgb}{0.2, 0.3, 0.1}

\usepackage[utf8]{inputenc}

\begin{document}

\title{A finite temperature framework for quark matter with color-superconducting phases}

\author{Hosein Gholami\,\orcidlink{0009-0003-3194-926X}}\email{mohammadhossein.gholami@tu-darmstadt.de}
\affiliation{Technische Universität Darmstadt, Fachbereich Physik, Institut f\"ur Kernphysik,
Theoriezentrum, Schlossgartenstr. 2, D-64289 Darmstadt, Germany}
\author{Marco Hofmann\,\orcidlink{0000-0002-4947-1693}}\email{marco.hofmann@tu-darmstadt.de (Corresponding Author)}
\affiliation{Technische Universität Darmstadt, Fachbereich Physik, Institut f\"ur Kernphysik, Theoriezentrum, Schlossgartenstr. 2, D-64289 Darmstadt, Germany}
\author{D\'ebora Mroczek\,\orcidlink{0000-0002-5417-6189}}\email{deboram2@illinois.edu}
\affiliation{Illinois Center for Advanced Studies of the Universe, Department of Physics, University of Illinois at Urbana-Champaign, Urbana, IL 61801, USA}
\author{Jacquelyn Noronha-Hostler\,\orcidlink{0000-0003-3229-4958}}
\affiliation{Illinois Center for Advanced Studies of the Universe, Department of Physics, University of Illinois at Urbana-Champaign, Urbana, IL 61801, USA}

\date{\today}

\begin{abstract}
\noindent
    Current observations of neutron stars and measurements of gravitational waves only provide constraints on the zero temperature ($T=0$) equation of state (EoS) of dense matter. The detection of the post-merger gravitational-wave signal from a binary neutron star merger would additionally provide access to finite-temperature properties of the EoS which contain more information about the composition and the interactions of dense matter than the cold EoS alone. In particular deconfined quark matter may be probed by its characteristic finite temperature effects. This is especially the case for color-superconducting phases, in which the quasiparticle contribution to the thermal pressure is exponentially suppressed at low temperatures. Here we develop a new finite $T$ 
    framework to model the thermal EoS for dense quark matter based on the cold quark matter EoS which is useful for numerical relativity simulations. We test the validity of the framework against a three-flavor NJL mean-field calculation, both with and without diquark pairing. We find that even for the complicated phase diagram of the NJL model including multiple different phases the framework is accurate to the few percent level for temperatures up to $T\sim 50$\,MeV.
\end{abstract}

\maketitle

\section{Introduction}

Neutron stars are stable astrophysical objects which reach the highest densities known in the universe, up to several times the nuclear saturation density ($n_{\rm sat} \simeq 0.16$\,fm$^{-3}$) in the core of the most massive stars recorded \cite{MUSES:2023hyz,Miller:2021qha,Landry:2020vaw,Mroczek:2023zxo}. 
Since the ground state of nuclear matter cannot be calculated from first-principles in the regime of quantum chromodynamics (QCD) realized in the core of neutron stars due to the fermion sign problem, we rely on observations to determine the relevant degrees of freedom and interactions. As a result, many models for new phases of matter have been proposed which are compatible with astrophysical observations (see Ref.~\cite{Mroczek:2023zxo}, Table I, for a review). 
Among these possibilities is that the densities in neutron stars are sufficiently high for quarks to become deconfined. 
At very high densities, where QCD becomes weakly coupled, the ground state of matter is a color-superconductor \cite{Son:1998uk,Schafer:1999jg,Pisarski:1999tv}, characterized by Cooper pairing between quarks due to attractive interactions. More specifically, three-flavor quark matter forms a color-flavor locked (CFL) phase \cite{Alford:1998mk,Schafer:1999fe,Shovkovy:1999mr} at large densities. At lower densities the two-flavor superconducting (2SC) phase \cite{Alford:1999pa} with only up and down pairing of two colors could be the favored state, due to pairing stress from the larger effective strange quark mass \cite{Alford:1997zt}. For reviews on color superconductivity (CSC), see Refs.~\cite{Buballa:2003qv,Shovkovy:2004me,Alford:2007xm,Schmitt:2025cqi}. Dense matter in neutron stars is not weakly coupled, still color-superconducting matter in the 2SC or CFL phase could be realized in the core of massive neutron stars or in binary neutron star mergers, see e.g. Refs.~\cite{Baldo:2002ju,Buballa:2003et,Klahn:2006iw,Pagliara:2007ph,Ranea-Sandoval:2017ort,Ivanytskyi:2022oxv,Ivanytskyi:2022bjc,Gartlein:2023vif,Gartlein:2024cbj,Gholami:2024ety,Christian:2025dhe}.

Information about the thermodynamic properties of a system, including signatures associated with the onset of new degrees of freedom and interactions is encoded in the equation of state (EoS). In cold, stable neutron stars, the EoS is the variable input which uniquely determines macroscopic properties such as the mass-radius sequence and the tidal deformability. 

Current constraints on cold neutron stars include macroscopic properties that were extracted from pulsar-timing observations by the Neutron Star Interior Composition ExploreR (NICER) collaboration \cite{Miller:2019cac,Miller:2021qha,Miller:2025qfq,Riley:2019yda,Riley:2021pdl,Choudhury:2024xbk} and gravitational-wave data from the inspiral between merging neutron stars by the LIGO/Virgo/KAGRA (LVK) collaboration \cite{LIGOScientific:2017vwq,De:2018uhw,LIGOScientific:2018cki,LIGOScientific:2020aai}. The existence of quarks in the core of cold, stable neutron stars is still an open question \cite{Alford:2004pf,Alford:2013aca,Baym:2017whm,McLerran:2018hbz,Tews:2018kmu,Annala:2019puf,Annala:2021gom,Annala:2023cwx,Kurkela:2014vha,Tan:2021nat,Tan:2021ahl,Legred:2021hdx,Gorda:2022jvk,Mroczek:2023zxo,Kurkela:2024xfh}. Beyond that, identifying specific types of quark phases in cold, stable neutron stars is currently beyond our reach, although a number of works have focused on color superconducting phases in particular \cite{Baldo:2002ju,Buballa:2003et,Klahn:2006iw,Pagliara:2007ph,Ranea-Sandoval:2017ort,Ivanytskyi:2022oxv,Ivanytskyi:2022bjc,Gartlein:2023vif,Gartlein:2024cbj,Gholami:2024ety,Christian:2025dhe}.

A challenge in detecting possible quark phases is that the macroscopic properties of cold neutron stars that we can observe may not be strongly sensitive to quark matter, such that hybrid neutron stars with a quark core may masquerade as a neutron star without a quark core due to nearly identical macroscopic signatures \cite{Alford:2004pf}.
Therefore, the detection of dense quark matter may require observables that are sensitive to finite temperature ($T$) effects.
Color-superconducting quark matter exhibits characteristic thermal effects -- due to the diquark gap in the quasiparticle spectrum, the thermal pressure contribution of the qausiparticles is suppressed exponentially at $T$ much smaller than the critical temperature $T_c$ at which the diquark condensates melt \cite{Fukushima:2004zq}.

The finite temperature EoS is probed in binary neutron star mergers, where current simulations estimate that the system can reach up to about $T \approx 50$ MeV \cite{Most:2018eaw,Most:2019onn,Most:2021ktk,Fields:2023bhs} (although higher temperatures may be possible in certain conditions).
Future observing runs at LVK  or next-generation gravitational wave detectors may be able to add extra information about the post-merger stage of binary neutron star mergers \cite{Torres-Rivas:2018svp} that may be sensitive to these finite $T$ effects. It is therefore important to study the imprints of different dense matter phases on observables such as the gravitational waveform from binary neutron star mergers using numerical relativity simulations.

Numerical relativity simulations require a crust-to-core thermodynamic description of neutron stars at finite temperature, meaning that the EoS must be tabulated for a wide range of densities at temperatures up to $T\approx 100$\,MeV.
A significant challenge is the computational cost of generating these EoS tables from microscopic calculations for numerical implementation.
Solutions proposed in the literature include ideal fermi gas approximations \cite{Janka:1993} or expansions which make assumptions about the degrees of freedom (i.e.~neutrons, protons, and electrons only \cite{Raithel:2019gws,Raithel:2021hye,Most:2021ktk}), which are not suitable for 2SC or CFL quark matter. Recently, an extension of the cold EoS of unpaired quark matter to finite temperatures and arbitrary electron fractions based on Fermi liquid theory was proposed and tested against an MIT bag model \cite{Zhu:2025vmz}.

An alternative was presented in Ref.~\cite{Mroczek:2024sfp}, which treats the thermal component of the pressure as a Taylor series in $(T/\mu_B)$, where $\mu_B$ is the baryon chemical potential. 
This approach allows the finite temperature EoS to be reconstructed from coefficients related to the entropy of the system as the temperature approaches zero. 
These coefficients can be parametrized or directly calculated from microscopic models to reproduce a particular EoS. 
Ref.~\cite{Mroczek:2024sfp} demonstrated that this formalism works well for the mean-field nucleonic model from Ref.~\cite{Alford:2022bpp}. 

In this work, we expand upon Ref.~\cite{Mroczek:2024sfp} by building a framework that can approximate the finite temperature EoS in both the 2SC and CFL phases and enforces thermodynamic stability. 
We test our framework against the three-flavor Nambu--Jona-Lasinio (NJL) model in the mean-field approximation in the renormalization group consistent treatment developed in Ref.~\cite{Gholami:2024diy}. The NJL model allows to self-consistently calculate effective ("constituent") quark masses from chiral symmetry breaking as well as diquark pairing in one model and has been used extensively to calculate phase diagrams \cite{Ruester:2005jc,Ruester:2005ib,Abuki:2005ms,Blaschke:2005uj,Sandin:2007zr,Gholami:2024diy} and model hybrid stars including color-superconducting quark matter at zero temperature in the past \cite{Bonanno:2011ch,Klahn:2013kga,Kojo:2014rca,Baym:2017whm,Baym:2019iky,Tanimoto:2019tsl,Kojo:2021wax,Alaverdyan:2020xnv,Alaverdyan:2022foz, Gao:2024lzu,Gholami:2024ety,Christian:2025dhe}.
The Taylor series of Ref.~\cite{Mroczek:2024sfp} alone cannot describe the finite $T$ behavior in the model correctly due to quasiparticle modes in color-superconducting phases. Thus, we propose an
analytic approximation of the contribution of paired quarks to the pressure at finite $T$, which is valid for temperatures $T\ll T_c$. For higher temperatures, we enforce thermodynamic stability by requiring the positivity of the heat capacity.

Our new finite $T$ framework is a significant improvement to other commonly used approaches, which cannot accurately represent the complicated phase structure which appears in NJL models. We show that we can accurately describe the pressure with less than $\lesssim 5\%$ relative error up to $T\sim 50$ MeV and less than $\lesssim 20\%$ relative error up to $T\sim 100$ MeV in densities relevant for the core of stable neutron stars and simulations of neutron star mergers.

This work is organized as follows: In \cref{sec:methods} we explain the three-flavor NJL model used as a test case for the $T$ expansion of three-flavor (color-superconducting) quark matter. We shortly review the $T$ expansion from Ref.~\cite{Mroczek:2024sfp} which we use for the pressure contribution from the unpaired quarks and introduce a $T$ expansion for paired quarks, again based on the cold EoS. In \cref{sec:results} we show our results for the phase diagram of the model, the terms in our expansion and the accuracy of the reconstructed EoS based on the cold EoS for the NJL model with and without diquark pairing. We also calculate the thermal index $\Gamma_{\text{th}}$, and show that a constant thermal index approximation cannot capture the complex phase structure of the NJL model. 
We end with conclusions and outlook to future work in \cref{sec:conclusions}.

We use natural units where $\hbar=c=k_B=1$.

\section{Methods}\label{sec:methods}

In this section, we explain the NJL model that we use for dense quark matter, review and formulate the finite $T$ expansion of Ref.~\cite{Mroczek:2024sfp} for this model, and introduce an analytic formula to capture the contribution of the paired quarks.

\subsection{Model details}\label{sec:model}

We describe deconfined quark matter with possible Cooper pairing in a three-flavor NJL model in the mean-field approximation with self-consistent treatment of quark masses (with the bare masses $m_u,m_d,m_s$ for up, down, strange quarks, respectively). 
Our Lagrangian includes different interactions that lead to the following desired properties:
\begin{itemize}
    \item chiral symmetry breaking from a scalar-pseudoscalar four-quark interaction with coupling \(G_S\) (mass dimension -2).
    \item \(U_A(1)\) breaking from  a six-quark Kobayashi-Maskawa-'t Hooft (KMT)  determinant \cite{Kobayashi:1970ji,tHooft:1976rip} with coupling \(K\) (mass dimension -5).
    \item spin-zero diquark pairing from a scalar diquark interaction in the color- and flavor antitriplet channel with coupling \(G_D\) (mass dimension -2).
\end{itemize}
In addition to the six parameters mentioned above:
$\{m_u, m_d, m_s, G_S, K, G_D\}$,
we have \(\Lambda'\), the renormalization-group (RG) scale at which the UV classical effective action is defined, i.e., the scale at which we specify the Lagrangian of the model.
Altogether, the model therefore contains seven free parameters,
$\mathbf{\Theta} = \{m_u, m_d, m_s, G_S, K, G_D, \Lambda'\}.$
Typically, six of these are fixed by requiring the model to reproduce vacuum meson observables, while \(G_D\) is treated as an unconstrained free parameter. 

 The Lagrangian reads \cite{Rehberg:1995kh,Gastineau:2001zke}
\begin{widetext}
\begin{align}
\mathcal{L}&=\bar{\psi}\bigl(i\slashed{\partial}+\gamma^0\hat{\mu}-\hat{m}\bigr)\psi
+G_S\sum_{A=0}^{8}\Big[(\bar{\psi}\tau_A\psi)^2+(\bar{\psi}i\gamma_5\tau_A\psi)^2\Big]
 -K\Big[\det_{\!f}\bigl(\bar{\psi}(\mathds{1}+\gamma_5)\psi\bigr)+\det_{\!f}\bigl(\bar{\psi}(\mathds{1}-\gamma_5)\psi\bigr)\Big]
\nonumber\\
&+  G_D
\big[\,\big(\bar{\psi}^a_\alpha i\gamma_5 \epsilon^{\alpha\beta\gamma} \epsilon_{abc}(\psi_C)^b_\beta\big)
\big((\bar{\psi}_C)_r^\rho i\gamma_5 \epsilon_{\rho\sigma\gamma} \epsilon^{rsc} \psi_s^\sigma\big)+\; \big(\bar{\psi}^a_\alpha  \epsilon^{\alpha\beta\gamma} \epsilon_{abc}(\psi_C)^b_\beta\big)
\big((\bar{\psi}_C)_r^\rho  \epsilon_{\rho\sigma\gamma} \epsilon^{rsc} \psi_s^\sigma\big) \,\big]\nonumber,
\label{eq:njl_L}
\end{align}
\end{widetext}
where the quark fields $(\psi)$ carry flavor \(\alpha=u,d,s\) and color \(a=r,g,b\) quantum numbers, with current quark masses \(\hat{m}=\mathrm{diag}_f(m_u,m_d,m_s)\). The \(\tau_A\) are Gell-Mann matrices in flavor space with \(\tau_0=\sqrt{2/3}\,\mathds{1}_f\), and \(\psi_C\equiv C\bar{\psi}^T\) with \(C=i\gamma^2\gamma^0\).

We solve the model in the mean-field approximation, allowing for chiral condensates \(\phi_\alpha=\langle\bar{\psi}_\alpha\psi_\alpha\rangle\) (\(\alpha=u,d,s\)) and scalar diquark condensates
\(\Delta_A\) (\(A=1,2,3\)) in the color-flavor antitriplet channel,
\begin{align}
\Delta_A=-2G_D\langle \bar{\psi}^a_\alpha i\gamma_5 \epsilon^{\alpha\beta A}\epsilon_{abA}(\psi_C)^b_\beta\rangle,
\end{align}
which distinguish two-flavor color superconductivity (2SC) (\(\Delta_3\neq0,\;\Delta_{1,2}=0\)) and color flavor locked (CFL) (\(\Delta_1,\Delta_2,\Delta_3\neq0\)) phases.

The chemical potential matrix ($\hat{\mu}$) in \cref{eq:njl_L} is given by
\begin{align}
\hat{\mu}^{\alpha\beta}_{ab}=&\bigl(\frac{1}{3}\mu_B\,\delta^{\alpha\beta}+\mu_Q\,Q^{\alpha\beta}+\mu_S S^{\alpha\beta} \bigr)\delta_{ab} \nonumber\\
&+\bigl(\mu_3(\lambda_3)_{ab}+\mu_8(\lambda_8)_{ab}\bigr)\delta^{\alpha\beta},
\label{eq:mu_matrix_main}
\end{align}
in color and flavor space, with the baryon chemical potential $\mu_B$, the generators of electric charge \(Q=\mathrm{diag}_f(2/3,-1/3,-1/3)\), strangeness \(S=\mathrm{diag}_f(0,0,-1)\) and the color generators (i.e., Gell-mann matrices in color space) \(\lambda_{3}\) and $\lambda_8$. The factor of 1/3 in front of $\mu_B$ appears because the quark number chemical potential is 1/3 of the baryon number chemical potential.

In this work, we focus on isospin-symmetric matter with \(\mu_Q=\mu_S=0\) for the electric charge and strangeness chemical potential, respectively, and leptons are not considered. In neutron star matter, $\mu_Q$ is determined from the beta equilibrium condition $\mu_d=\mu_s=\mu_u+\mu_e$ (assuming neutrinoless matter). However, for the large-density limit of three flavor quark matter $\mu_Q=0$ with deviations at lower densities due to the nonzero constituent strange quark mass \cite{Danhoni:2025qpn}.

The mean-field effective potential, $\Omega_{\text{eff}}(T,\mu_B,\mathbf{X})$, is a function of the temperature, the baryon chemical potential, and the condensates and the color chemical potentials summarized in $\mathbf{X}=\{\phi_u,\phi_d,\phi_s,\Delta_1,\Delta_2,\Delta_3,\mu_3,\mu_8 \}$.
The color-neutral physical solution is obtained by imposing that \(\Omega_\text{eff}\) is stationary with respect to all quantities in $\mathbf{X}$, i.e., by solving the gap equations,
    \begin{align}\label{eq:gapeq}
    \frac{\partial \Omega_\text{eff}}{\partial \phi_\alpha}\bigg|_{\phi_\alpha=\bar{\phi}_\alpha}=\frac{\partial \Omega_\text{eff}}{\partial \Delta_A}\bigg|_{\Delta_A=\bar{\Delta}_A}=0\,,
\end{align}
for $\alpha=u,d,s$ and $A=1,2,3$ and the color neutrality conditions,
\begin{equation}\label{eq:neutrality}
\frac{\partial \Omega_\text{eff}}{\partial \mu_3}\bigg|_{\mu_3=\bar{\mu}_3}=\frac{\partial \Omega_\text{eff}}{\partial \mu_8}\bigg|_{\mu_8=\bar{\mu}_8}=0\,,
\end{equation}
self-consistently. If multiple solutions exist, we select the one with the smaller $\Omega_{\text{eff}}$. The corresponding mean fields and auxiliary chemical potentials are denoted $\mathbf{\bar{X}}=\{\bar{\phi}_u,\bar{\phi}_d,\bar{\phi}_s,\bar{\Delta}_1,\bar{\Delta}_2,\bar{\Delta}_3,\bar{\mu}_3,\bar{\mu}_8 \}$. Inserting these solutions back into the effective potential gives the thermodynamic potential,
\begin{align}\label{eq:deffomega}
    \Omega_\text{}(T,\mu_B)\equiv\Omega_\text{eff}\big(T,\mu_B,\mathbf{\bar{X}}\big).
\end{align}
The pressure is then normalized to zero in vacuum:
\begin{align}
P(T,\mu_B)
= -\Omega_\text{}(T,\mu_B)
+\Omega_\text{}(0,0).
\end{align}
All other thermodynamic quantities follow from the usual thermodynamic relations, e.g. the entropy density
\begin{eqnarray}
s&=&\frac{\partial P}{\partial T}\Big|_{\mu_B,\mu_Q,\mu_S}.
\end{eqnarray}

Since the NJL model is nonrenormalizable, it must be regularized with a finite energy scale. The conventional choice is a sharp three-momentum cutoff, which can introduce cutoff artifacts, especially in the presence of diquark condensates at high density~\cite{Farias:2005cr}. To avoid such artifacts, we evaluate the mean-field effective potential using the renormalization-group consistent (RG-consistent) treatment~\cite{Braun:2018svj,Gholami:2024diy} and, in particular, employ its minimal scheme variant, which is motivated by the wave-function renormalization of the diquark field~\cite{Gholami:2024diy,Gholami:2025afm}, and is consistent with the Bardeen–Cooper–Schrieffer (BCS) relation for the critical temperature~\cite{Gholami:2025guq}.

We summarize the different phases of quark matter that may appear in the three-flavor NJL model and what their properties are in \cref{tab:phasesofmatter}. 
The chiral symmetry breaking
($\chi$SB) phase is characterized by the absence of diquark pairing and by
large light-quark chiral condensates, $\langle \bar{\psi}_u \psi_u \rangle,\; \langle \bar{\psi}_d \psi_d \rangle$. In the Normal Quark Matter (NQM) phase,
diquark pairing is also absent, but the chiral condensates of the light flavors are significantly
reduced, restoring chiral symmetry in the light flavors approximately. The strange quark fraction $Y_s$ varies most strongly at large
$\mu_B$ and $T$, where the effective strange-quark mass decreases. The 2SC phase features diquark pairing between up and down
quarks of two colors (e.g. red and green) with the third color (blue) unpaired. In this phase $n_u = n_d\neq n_s$ and
$Y_s$ continues to evolve as the system moves to higher $T$ and $\mu_B$ due to a decrease of the strange quark mass.  
Finally, the CFL phase exhibits pairing across all three
flavors, leading to approximately equal number densities for the light and
strange quarks, $n_u = n_d \approx n_s$. The strange quark fraction changes only
weakly in the CFL phase, as the strange quark mass is already relatively small compared to the quark chemical potential.
\begin{table*}[t]
    \centering
    \begin{tblr}{
        width=\textwidth,
        colspec={X[c] X[c,0.6] X[c,0.6] X[c]},
        row{1}={font=\bfseries},
        hline{1,Z}={1pt},
        hline{2}={0.5pt}
    }
        Phase & Acronym & Diquark pairing & Properties in symmetric matter \\
        Chiral Symmetry Breaking & $\chi$SB & No & large $\langle \bar{\psi}_u \psi_u \rangle,\; \langle \bar{\psi}_d \psi_d \rangle$ \\
        Normal Quark Matter & NQM & No & small $\langle \bar{\psi}_u \psi_u \rangle$ and $ \langle \bar{\psi}_d \psi_d \rangle$; $Y_s$ varies \\
        Two-flavor Color Superconducting & 2SC & Yes & $n_u = n_d$; $Y_s$ varies \\
        Color Flavor Locked & CFL & Yes & $n_u = n_d \approx n_s$ \\
    \end{tblr}
    \caption{Phases of quark matter in our 2+1 flavor NJL model and relevant properties.}
    \label{tab:phasesofmatter}
\end{table*}

\subsection{Finite \texorpdfstring{$T$}{T} expansion for unpaired quarks}\label{sec:unpaired_expansion}

In Ref.~\cite{Mroczek:2024sfp}, a Taylor series was used to expand the $T=0$ pressure up to $T\sim 100$ MeV for both isospin-symmetric and asymmetric matter. 
Here, we will write that expansion with a focus on isospin-symmetric matter,
where the electric charge chemical potential is $\mu_Q=\mu_S=0$ (so we only have one chemical potential, the baryon chemical potential $\mu_B$, to track). 
The pressure expansion appears as:
\begin{align}\label{eq:thermos0}
\dfrac{1}{T^4}P^{\text{Taylor}}(T, \mu_B)=& \dfrac{1}{T^4}P(T=0,\mu_B)+\tilde{c}_1(\mu_B)\left(\dfrac{T}{\mu_B}\right)\nonumber\\
&+\frac{1}{2} \tilde{c}_2 (\mu_B)\left(\dfrac{T}{\mu_B} \right)^2\nonumber\\
&+\frac{1}{6}\tilde{c}_3(\mu_B) \left(\dfrac{T}{\mu_B} \right)^3\nonumber\\&+\frac{1}{24}\tilde{c}_4(\mu_B)\left(\dfrac{T}{\mu_B} \right)^4\nonumber+\mathcal{O}\left(\dfrac{T}{\mu_B} \right)^5,
\end{align}
with the temperature expansion coefficients $\tilde{c}_i(\mu_B)$ defined as:
\begin{align}
\tilde{c}_i(\mu_B)=& \frac{\partial^i (P/T^4)}{\partial (T/\mu_B)^i}\biggr\rvert_{\mu_B}(T=0,\mu_B).
\end{align}
Since the expansion is centered around $T=0$, we can write,

\begin{align}\label{eq:thermos}
P^{\text{Taylor}}(T, \mu_B)=& P(T=0,\mu_B)+c_1(\mu_B)T\nonumber\\
&+\frac{1}{2} c_2 (\mu_B)T^2\nonumber+\frac{1}{6}c_3(\mu_B)T^3\nonumber\\&\frac{1}{24}c_4(\mu_B)T^4+\mathcal{O}(T)^5,
\end{align}

with the coefficients $c_i(\mu_B)$ now defined as:
\begin{align}
c_i(\mu_B)=& \frac{\partial^i P}{\partial T^i}\biggr\rvert_{\mu_B}(T=0,\mu_B).
\end{align}

Specifically, we can write the coefficients as derivatives of the entropy density $s(T, \mu_B)=\partial P/\partial T\rvert_{\mu_B}$ and write the expansion in \cref{eq:thermos} as
\begin{align}\label{eq:T_expansion_normal}
P^{\text{Taylor}}(T, \mu_B)=& P(T=0,\mu_B)+\frac{1}{2}\frac{\partial s}{\partial T}\biggr\rvert_{\mu_B}(T=0,\mu_B)\cdot T^2\nonumber\\&+\frac{1}{6}\frac{\partial^2 s}{\partial T^2}\biggr\rvert_{\mu_B}(T=0,\mu_B)\cdot T^3\nonumber\\&
+\frac{1}{24}\frac{\partial^3 s}{\partial T^4}\biggr\rvert_{\mu_B}(T=0,\mu_B)\cdot T^3\nonumber\\
&+\mathcal{O}(T^5),
\end{align}
where the linear term $s(T=0)\cdot T$  vanishes in all standard neutron star EoS models. For a relativistic mean-field model with nucleons, it was found that terms up to order $T^2$ were sufficient to describe the pressure accurately up to $T\sim 100$ MeV (especially at large $\mu_B$) \cite{Mroczek:2024sfp}.

The pressure in our mean-field model depends on the mean-field values of the condensates $\mathbf{\bar{X}}(T,\mu_B)$, see \cref{eq:deffomega}, which have a non-trivial temperature dependence from the solution of the gap equations \cref{eq:gapeq} and neutrality conditions \cref{eq:neutrality}. Thus, the coefficients in the $T$ expansion should be calculated
via a constrained derivative (imposing the stationarity of the condensates through the gap equations and neutrality conditions) of the effective potential \(\Omega_\text{eff}\).
In Ref.~\cite{Gholami:2025cfq}, it was shown that these constrained derivatives admit a natural expansion around the derivative taken \emph{without} constraints (the “naive” derivative). Denoting \(H_{ij}\equiv \partial^2\Omega_\text{eff}/\partial X_i\partial X_j\) as the Hessian in field space, one finds schematically,
\begin{align}
\left.\frac{\mathrm{d}\Omega_\text{eff}}{\mathrm{d}T}\right|_{\text{constr.}}
&= \left.\partial_T \Omega_\text{eff}\right|_{\text{naive}},\\
\left.\frac{\mathrm{d}^2\Omega_\text{eff}}{\mathrm{d}T^2}\right|_{\text{constr.}}
&= \left.\partial_T^2 \Omega_\text{eff}\right|_{\text{naive}}
\;-\; \sum_{i,j} \Omega_{T i}\,(H^{-1})_{ij}\,\Omega_{T j}
\;,\\
\left.\frac{\mathrm{d}^i\Omega_\text{eff}}{\mathrm{d}T^i}\right|_{\text{constr.}}
&= \left.\partial_T^i \Omega_\text{eff}\right|_{\text{naive}}
+\;\mathcal{J},
\end{align}
where $\mathcal{J}$ denotes higher-order terms involving increasing powers of \(H^{-1}\), and \(\Omega_{T i}\equiv \partial_T\partial_{X_i}\Omega_\text{eff}\). These cannot be written in a compact form but are obtained recursively from lower-order constrained derivatives.
Ref.~\cite{Gholami:2025cfq} demonstrated that, away from phase transitions along the temperature axis, the naive derivative is sufficient to approximate the constrained derivatives at low temperatures.
We adopt the same reasoning here, as we are only interested in $T$ derivatives at zero temperature. Thus, we calculate the coefficients in \cref{eq:thermos} using naive derivatives:
\begin{align}
c_i(\mu_B)=& \frac{\partial^i P}{\partial T^i}\biggr\rvert_{\mu_B}(T=0,\mu_B)\\\approx&\frac{\partial^i \Omega_\text{eff}}{\partial T^i}\biggr\rvert_{\mu_B}(T=0,\mu_B,
\mathbf{\bar{X}}(T=0),\mu_B).
\end{align}
Using the above equation, we can write \cref{eq:T_expansion_normal} entirely in terms of partial derivatives of the effective potential with respect to $\mu_B$.

\subsection{Finite \texorpdfstring{$T$}{T} expansion for paired quarks}\label{sec:paired_expansion}
 The $T$ expansion in \cref{eq:T_expansion_normal} of the previous section is a Taylor expansion in $T$, i.e. it requires that the pressure is an analytic function in the temperature. This is not the case anymore for the pressure contribution of Cooper paired quarks. This is due to the fact that the coefficients in \cref{eq:thermos} are calculated at zero temperature, where the contribution of the paired modes to the pressure and to all derivatives of the pressure is non-analytic, meaning that all coefficients $c_i(\mu_B)$ are vanishing. We therefore introduce a new $T$ expansion for paired quarks.
 
We estimate in \cref{app:paired_contr} the pressure of a paired (quasiparticle) mode with the dispersion relation, 
\begin{equation*}
    \epsilon=\sqrt{(p-\bar{\mu})^2+\Delta^2}\pm \delta\mu/2.
\end{equation*}
This is the dispersion relation of a massless quasiparticle with momentum $p$, diquark gap $\Delta$, the average chemical potential $\bar{\mu}=(\mu_1+\mu_2)/2$ and chemical potential mismatch $\delta\mu=(\mu_1-\mu_2)/2$ of the paired quark species with chemical potentials $\mu_1$ and $\mu_2$, respectively. We separate the pressure $P^{\text{single}}$ provided by a single quasiparticle into a cold part at zero temperature and into a thermal part:
\begin{align}
    P^{\text{single}}=P(T=0,\mu_B)
    + P_{\text{th}}^{\text{single}}(T,\mu_B).
\end{align}
The thermal pressure contribution can be approximated for large $\Delta/T$ and large $\bar{\mu}/T$ (degenerate matter) to (see \cref{app:paired_contr}):
\begin{align}\label{eq:paired_single}
     P_{\text{th}}^{\text{single}}(T,\mu_B)=&
\frac{\bar{\mu}^2T^2}{\pi^{3/2}}\sqrt{\Delta/T}\cdot \exp{\left(-\frac{\Delta}{T}\right)}\nonumber\\&\cdot\left(\sqrt{2}-\frac{1}{2}\exp{\left(-\frac{\Delta}{T}\right)}\right).
\end{align}
We use directly the formula derived for the zero Fermi momentum mismatch $\delta\mu=0$, as this is the case for isospin-symmetric matter at zero temperature. 

The expression in \cref{eq:paired_single} is an approximation for the pressure of a single quasiparticle. Depending on the color-superconducting phase, different numbers of quasiparticles with different diquark gaps, $\Delta$, and average chemical potential, $\bar{\mu}$, appear in the expression of the temperature. To obtain the full finite $T$ expansion for a color-superconducting phase, we add the sum of the pressure contributions of all quasiparticles with their degeneracy factor $g_i$:
\begin{align}\label{eq:P_Delta}
   P^\Delta(T,\mu_B)=& \sum_{i \in \text{pairings}}
g_i\frac{\bar{\mu}_i^2T^2}{\pi^{3/2}}\sqrt{\Delta_i/T}\cdot \exp{\left(-\frac{\Delta_i}{T}\right)}\nonumber\\&\qquad\cdot\left(\sqrt{2}-\frac{1}{2}\exp{\left(-\frac{\Delta_i}{T}\right)}\right).
\end{align}

We summarize the values of $g_i$, $\Delta_i$ and $\bar{\mu}_i$ for the 2SC and the CFL phase in \cref{tab:T_exp} and explain their values in the following:
\begin{itemize}
\item In the 2SC phase, red up quarks pair with green down quarks and green up quarks pair with red down quarks while the blue quarks remain unpaired. In total, we have $g_{\text{2SC,paired}}=4$ quasiparticle modes with the average chemical potential $\bar{\mu}_{\text{2SC,paired}}=\bar{\mu}_{ur,dg}=\frac{\mu_{ur}+\mu_{dg}}{2}=\frac{\mu_B}{3}+\frac{\mu_Q}{6}+\frac{\mu_8}{\sqrt{3}}$
and the gap $\Delta_3$. Note that in the 2SC phase, there are also unpaired blue quarks and (depending on the density) unpaired strange quarks of all three colors. They give nonzero contributions to the ordinary Taylor expansion $P^{\text{Taylor}}(T,\mu_B)$ in \cref{eq:T_expansion_normal}.
\item In the CFL phase, all quarks are paired. The massless, flavor symmetric CFL phase contains an octet of $g_{\text{octet}}=8$ modes with gap $\Delta_3=\Delta_2=\Delta_1$ and a singlet ($g_{\text{singlet}}=1$) with gap $2\Delta_3$. Adding both contributions gives the total pressure approximation.
Note that \cref{eq:P_Delta} was derived for massless quarks. In \cref{sec:coefficients} and \cref{sec:reconstruction}, we test the new $T$ expansion against the numerical NJL mean-field calculation, in which quark masses were calculated self-consistently, such that $\Delta_3>\Delta_1=\Delta_2$ holds in the CFL phase. For simplicity, we use the octet-singlet decomposition with gaps $\Delta_3$ and $2\Delta_3$ nonetheless.
\end{itemize}

\begin{table*}[t]
    \centering
    \begin{tblr}{
        width=\textwidth,
        colspec={X[c] X[c] X[c] X[c] X[c]},
        row{1}={font=\bfseries},
        hline{1,Z}={1pt},
        hline{2}={0.5pt}
    }
        phase & pairing pattern & $g_i$ & $\bar{\mu}_i$ & $\Delta_i$\\
        2SC & $(u_r,d_g),(u_g,d_r)$ & 4 & $\bar{\mu}_{ur,dg}=\frac{\mu_B}{3}+\frac{\mu_Q}{6}+\frac{\mu_8}{\sqrt{3}}$ & $\Delta_3$\\
        CFL & all $(i_a,j_b)$ with $i\neq j$, $a\neq b$ & 8\,(octet), 1\,(singlet) & $\frac{\mu_B}{3}$ & $\Delta_3$ (octet), $2\Delta_3$ (singlet)\\
    \end{tblr}
    \caption{Flavor--color pairing patterns, number of quasiparticle modes $g_i$, average chemical potentials $\bar{\mu}_i$, and gaps $\Delta_i$ to be inserted into \cref{eq:general_pairings} for the finite-$T$ expansion for paired quarks.}
    \label{tab:T_exp}
\end{table*}
So far, we did not take any temperature dependence of the diquark gaps into account. The temperature dependence of the constituent quark masses and the diquark gaps results from solving the gap equations, and are in general not trivial. For the diquark gaps, however, the simple analytic formula based on BCS theory \cite{Muhlschlegel1959}
\begin{align}\label{eq:Delta_T_approx}
    \Delta(T)=\theta(T_c-T)\Delta(T=0)(1-(T/T_c)^{3.4})^{0.53}
\end{align}
can be used. The critical temperature for the phase transition from 2SC to NQM can be estimated from the diquark gap at zero temperature via \cite{Schmitt:2002sc}
\begin{equation}\label{eq:Tc_approx}
T_c^{\text{2SC$\to$ NQM}} \approx 
\left\{
\begin{array}{ll}
  T_c^{\text{BCS}}, & \text{if 2SC at } T=0, \\[6pt]
  2^{-1/3} T_c^{\text{BCS}}, & \text{if CFL at } T=0,
\end{array}
\right.
\end{equation}
with the BCS relation $T_c^{\text{BCS}}\approx 0.57 \Delta_3(T=0)$ \cite{Bardeen:1957mv}, which is exact for massless quarks~\cite{Gholami:2025afm,Gholami:2025guq}.
We show how this parameterization of the temperature dependence of the diquark gap compares to the calculation in the NJL model in \cref{app:Delta_t}.

With increasing temperature, matter in our model which is in the CFL phase at $T=0$ will first melt to a 2SC phase before the 2SC-NQM transition, see e.g. the phase diagrams in the bottom panels of \cref{fig:composition}, in which case the second formula in \cref{eq:Tc_approx} has to be applied \cite{Schmitt:2002sc}.
With the estimate \cref{eq:Tc_approx}, we can use \cref{eq:Delta_T_approx} to improve the pressure expansion below the critical temperature. For the CFL phase, we use the critical temperature of $\Delta_3$ for simplicity, which is higher than the critical temperature of $\Delta_1$ and $\Delta_2$ (see \cref{fig:delta_t}) due to the larger constituent strange quark mass, $M_s\sim 150\,$MeV, compared to the light up and down constituent quark masses $M_u,M_d\sim 20\,$MeV. In the limit of asymptotically large densities, the strange mass in our model approaches zero and the three diquark condensates and their critical temperatures converge to a constant value which depends on the diquark coupling \cite{Gholami:2025guq}. Consequently, we expect our expansion for the CFL phase to improve with increasing chemical potentials.

Already at a temperature $T^\ast <T_c$, the expression \cref{eq:P_Delta} with the temperature dependence \cref{eq:Delta_T_approx} inserted, becomes thermodynamically unstable, signaled by a negative $\partial^2 P ^\Delta/\partial T^2$ and eventually vanishing pressure, see \cref{app:extrapolation}. This is because $P^\Delta$ only describes the pressure of the quasiparticles, neglecting that their melting leads to unpaired quarks which provide additional thermal pressure. For our case, it is only relevant that the constructed thermal EoS shows no instabilities. Thus, instead of modeling the melting of the quasiparticles in detail, we calculate the temperature $T^\ast$ at which the instability occurs and linearly extrapolate $P^\Delta(T,\mu_B)$ with the slope given by the value of the entropy density $s(T^\ast,\mu_B)$ at $T^\ast$:
\begin{align}\label{eq:extrapolation}
    &P^{\text{quasip.}}(T,\mu_B)\nonumber\\
    &=\begin{cases}
        P^{\Delta}(T,\mu_B), \quad T< T^\ast\\
         P^\Delta(T^\ast,\mu_B) + s(T^\ast,\mu_B) (T-T^\ast) , \quad T> T^\ast.
    \end{cases}
\end{align}
Here, 
\begin{equation}
    T^\ast \simeq \begin{cases}
        0.9039\,T_c \quad (\text{2SC}) \\
        0.8889\,T_c \quad (\text{CFL}) 
    \end{cases}
\end{equation}
depends only on the critical temperature $T_c^{\text{2SC}\to\text{NQM}}$ from \cref{eq:Tc_approx} and  
\begin{equation}
    s(T^\ast,\mu_B) = \frac{\partial P^\Delta}{\partial T}\Bigg\vert_{T=T^\ast}
    \simeq \begin{cases}
        8.944  \dfrac{\bar{\mu}^2}{\pi^{3/2}}T^\ast \quad (\text{2SC}) \\\\
        12.09 \dfrac{\bar{\mu}^2}{\pi^{3/2}}T^\ast\quad (\text{CFL}) 
    \end{cases}\,,
\end{equation}
see \cref{app:extrapolation}. The final formula for calculating the pressure at nonzero temperatures is given by adding the pressure of the quasiparticles to the Taylor expansion for the unpaired quarks:
\begin{equation}\label{eq:general_pairings}
    P(T,\mu_B)=P^{\text{Taylor}}(T,\mu_B)+P^{\text{quasip.}}(T,\mu_B).
\end{equation}
In summary, our procedure for extending a cold EoS is the following: given $P(T=0,\mu_B)$ and the derivatives $c_n=\partial P/\partial T$ at zero temperature, together with the $ud$ pairing diquark gap $\Delta_3(T=0,\mu_B)$ and the chemical potentials of the paired quarks for every $\mu_B$ in the case of color-superconductivity, the EoS at nonzero temperatures can be approximated by \cref{eq:general_pairings} which is a Taylor polynomial of order $n$ plus an analytic expression for the thermal pressure of the quasiparticles. This expansion for color-superconducting phases is still an expansion of the zero temperature EoS: it uses only information at zero temperature in addition to the coefficients of the Taylor expansion \cref{eq:T_expansion_normal}.

\section{Results}\label{sec:results}

We test our proposed finite $T$ framework against the NJL model with the often-used RKH parameter set \cite{Rehberg:1995kh}, which fixes $\Lambda'=602.3$\,MeV, $G_S\,\Lambda'^2=1.835$, $K\,\Lambda'^5=12.36$ and the bare quark masses $m_{u,d}=5.5$\,MeV and $m_s=140.7$\,MeV. We show results both for the case of zero diquark coupling ($G_D=0$), i.e. no color-superconducting phases, and with the diquark coupling set to $G_D=G_S$, a typical value chosen to study the phase diagram of dense quark matter with different color-superconducting phases (see also Ref.~\cite{Ruester:2005jc,Gholami:2024diy,Gholami:2025afm} for phase diagrams in beta equilibrium).

In \cref{sec:pd}, we plot the phase diagrams together with the quark composition of the NJL model in the mean-field approximation. In \cref{sec:coefficients}, we then calculate the Taylor coefficients of the $T$ expansion and, in \cref{sec:reconstruction}, we reconstruct the finite $T$ EoS by applying our $T$ expansion. We test against the ``exact" NJL model calculation which solves the mean-field gap equations at finite $T$ with and without the new expansion for paired quarks. In \cref{sec:thermal_index}, we calculate the thermal index showing that is not a constant in unpaired quark matter and that it cannot be used for color-superconducting quark matter.
\subsection{Phase diagram}\label{sec:pd}
\begin{figure*}
    \begin{tabular}{cc}
      \includegraphics[width=0.5\linewidth]{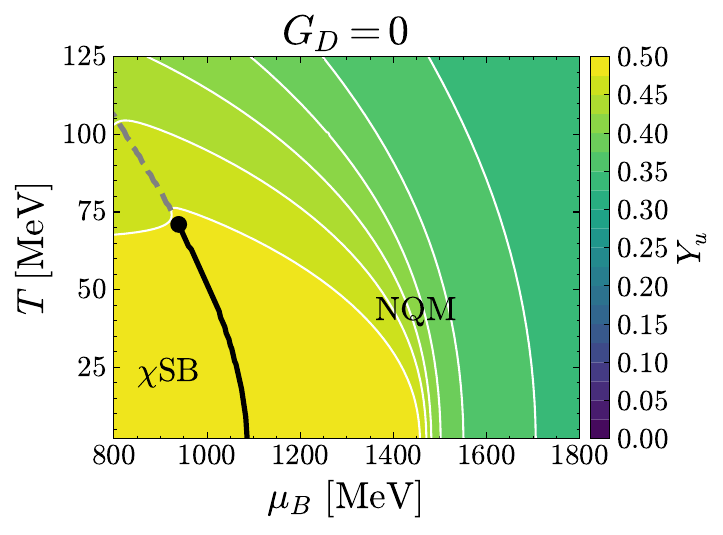}    &  \includegraphics[width=0.5\linewidth]{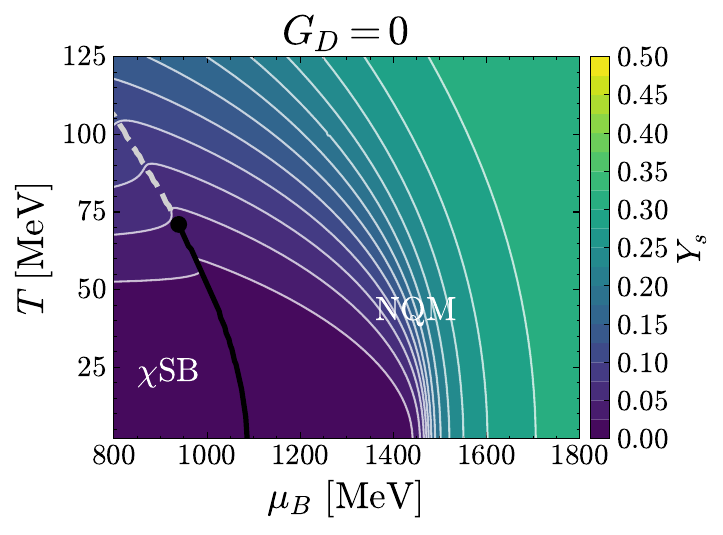}\\
         \includegraphics[width=0.5\linewidth]{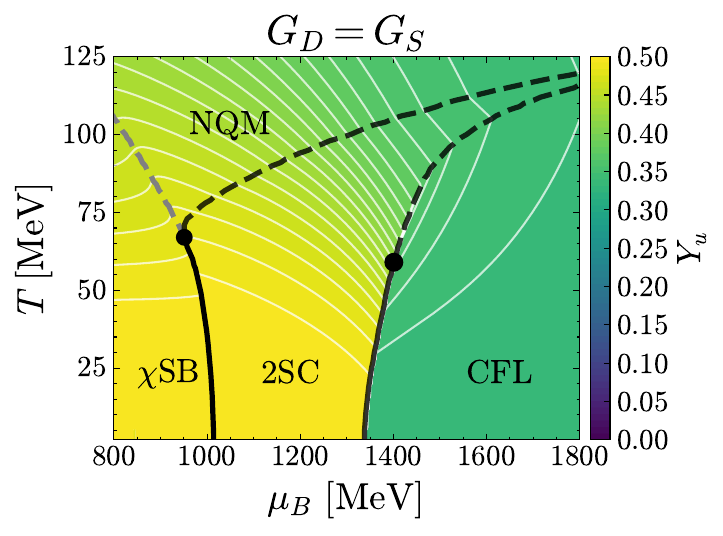}    &  \includegraphics[width=0.5\linewidth]{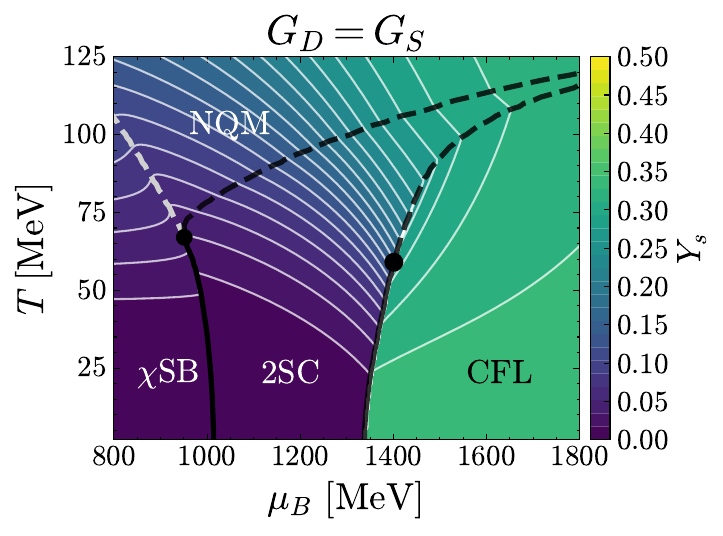}\\
    \end{tabular}
    \caption{Phase diagrams of the RG-consistent NJL model (minimal scheme) without a diquark coupling ($G_D=0$, top panels), and with color superconductivity ($G_D=G_S$, bottom panels). Phase boundaries are shown as solid (first order), dashed (second order) and gray, dashed (crossover) lines. A heat plot with contours (white lines) of constant quark number fractions $Y_u=Y_d$ (left) and $Y_s$ (right) of up and down and strange quarks, respectively, is overlaid. The acronyms for the phases denote chiral symmetry breaking ($\chi{\text{SB}}$), normal quark matter (NQM), two-flavor color-superconducting quark matter (2SC) and color-flavor locked matter (CFL), see \cref{tab:phasesofmatter}.}
    \label{fig:composition}
\end{figure*}

We show the phase structure of isospin-symmetric quark matter in the RG-consistent NJL model (minimal scheme) \cite{Gholami:2024diy,Gholami:2024ety,Christian:2025dhe} in \cref{fig:composition}. The top panels correspond to vanishing diquark coupling, \(G_D=0\) (no color superconductivity), while the bottom panels include diquark pairing, \(G_D=G_S\). In each panel, contours of the quark number fractions \(Y_u=Y_d\) and \(Y_s\) are overlaid on the phase boundaries in the \(T\)–\(\mu_B\) plane. 
The net quark number fractions for flavor $\alpha$ are calculated as 
\begin{equation}
    Y_\alpha=\frac{n_\alpha}{3n_B}=\frac{n_\alpha}{n_u+n_d+n_s}
\end{equation}
where $n_{\alpha}$ is the net quark number density of flavor $\alpha$ and the factor of 3 appears due to quarks carrying a baryon number of $B_\alpha=1/3$.  To be clear, we are studying here the quark fractions and not the baryon number fractions such that $Y_S=-3Y_s$ (note the lower-case $s$ for strange quarks instead of the upper case $S$ for strangeness in baryons), see Eq. (17) in Ref.~\cite{Danhoni:2025qpn} for comparison. 

First-order transitions are indicated by solid black lines, second-order transitions by dashed black lines, and the $\chi \text{SB}$-NQM crossover by a dashed gray line; black dots mark critical end points.
For the case where \(G_D=0\), the system is in the chiral-symmetry–broken phase \(\chi{\mathrm{SB}}\) at low \(T\) and \(\mu_B\). With increasing chemical potential or temperature chiral symmetry becomes restored approximately, which we denote as NQM. At low temperatures, this transition is first order. The first-order line terminates in a critical end point, above which the \(\chi{\mathrm{SB}}\)–NQM transition becomes a crossover. At high \(\mu_B\), the strange quark mass drops at a second crossover, most pronounced as the onset of strange quarks at zero temperature where the contour lines are dense. Consistent with the onset of strange quarks, the number fractions evolve from \(Y_u\simeq 0.5\) and \(Y_s\simeq 0\) at low \(T,\mu_B\) toward \(Y_u\simeq Y_s\simeq 1/3\) at higher \(T,\mu_B\).

With color superconductivity (bottom panels; \(G_D=G_S\)), the \(\chi{\mathrm{SB}}\) phase undergoes a first-order transition to 2SC at low \(T\) with increasing \(\mu_B\). The 2SC–CFL transition is first order at low \(T\) and terminates at a critical end point. Above it, the transition is second order. The \(\chi{\mathrm{SB}}\)–2SC line likewise ends at a critical point. At higher \(T\), the \(\chi{\mathrm{SB}}\)–NQM transition is a crossover and the 2SC–NQM boundary is second order. The RG-consistent treatment avoids the suppression of diquark pairing from the momentum cutoff in the conventional regularization of the model, which otherwise leads to the artifact of downward-bending of the color-superconducting phase boundaries \cite{Gholami:2024diy}. In our model, the CSC phase boundaries increase in temperature with \(\mu_B\) and converge to the CFL limit with zero quark masses at asymptotically large \(\mu_B\) \cite{Gholami:2025guq}.

The quark number fractions track the phase changes: at low \(T,\mu_B\), \(Y_u\simeq 0.5\) and remains near this value throughout 2SC. The strange-quark fraction is \(Y_s\simeq 0\) at low \(T,\mu_B\) and in 2SC. In the CFL phase, \(Y_u=Y_d=Y_s= 1/3\) at \(T=0\), with $Y_u$ increasing slightly above \(1/3\) at finite \(T\) within the color-superconducting region due to the larger strange mass. At asymptotically large $\mu_B$ or $T$ the fractions approach $Y_u=Y_d=Y_s=1/3$ both in color-superconducting and normal quark matter.

\subsection{Coefficients in the finite \texorpdfstring{$T$}{T} expansion}\label{sec:coefficients}

\begin{figure}
    \centering \includegraphics[width=\linewidth]{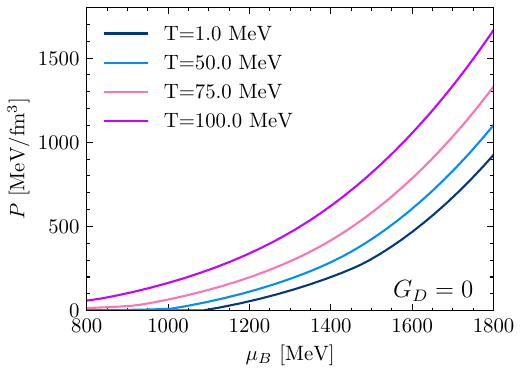}\\
    \centering \includegraphics[width=\linewidth]{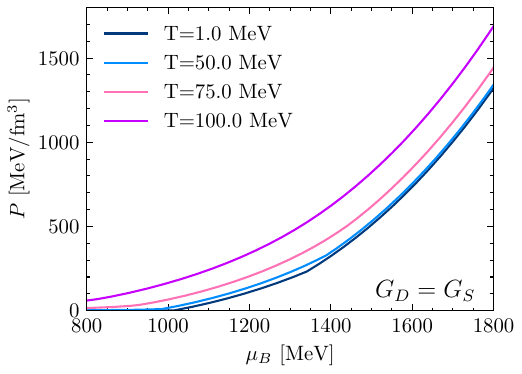} 
    \caption{The pressure versus the baryon chemical potential at different temperatures $T=1,50,75,100\,$MeV for the NJL model without diquark pairing (top) and with diquark pairing (bottom), respectively.}
    \label{fig:pressureT=0}
\end{figure}

\begin{figure}[htp]
    \centering
\begin{flushright}

\includegraphics[width=0.94\linewidth]{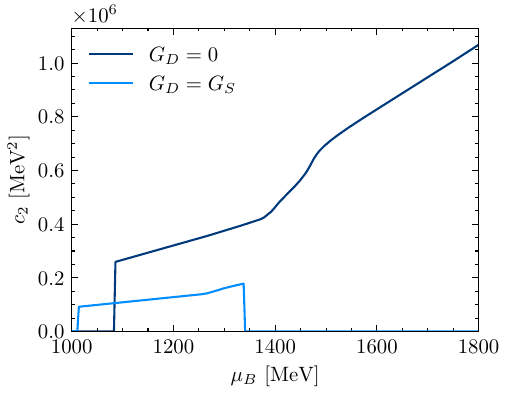}

\includegraphics[width=\linewidth]{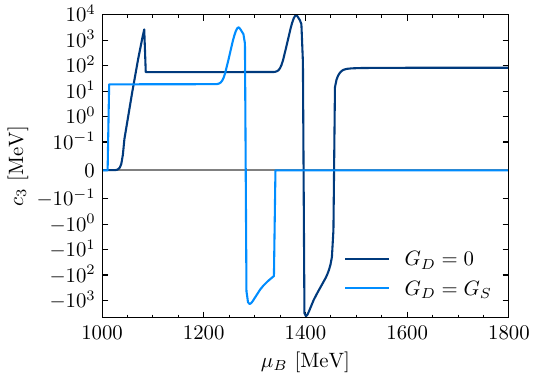}

\includegraphics[width=\linewidth]{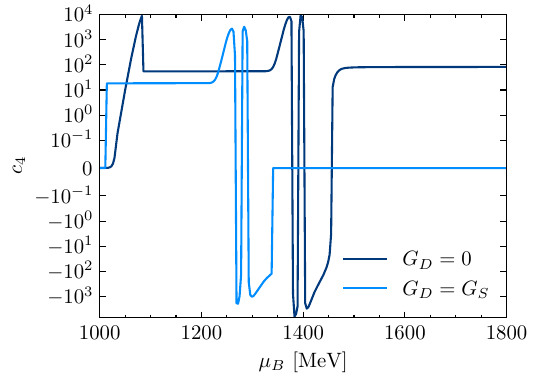}

\end{flushright}
    \caption{Derivatives $c_2=\partial s/ \partial T$ (top), $c_3=\partial^2 s/ \partial T^2$ and $c_4=\partial^3 s/ \partial T^3$ (symmetric-log plot, middle and bottom) which are coefficients for the $T^2$ term, the $T^3$ term and the $T^4$ term in the Taylor expansion, respectively. For easier numerics, the derivatives were calculated at $T=1\,$MeV. Both the model without diquark interactions ($G_D=0$) and with diquark interactions ($G_D=G_S$) are shown.}
    \label{fig:dsdt23_T=0}
\end{figure}

In this section, we calculate the Taylor expansion terms and discuss their impact on approximating the pressure at finite $T$.
We begin with the pressure at different temperatures vs baryon chemical potential in \cref{fig:pressureT=0}.
As one would expect from stability constraints, the pressure monotonically increases with $\mu_B$. 
We show the NJL model results both without ($G_D=0$) and with ($G_D=G_S$) diquark interactions. We can understand the results by comparing the phase diagrams in \cref{fig:composition}: 
if the ground state of the model is a color-superconducting phase, the pressure of the model with diquark pairing is larger than the pressure in the model without diquark pairing. If the ground-state of the model is normal conducting quark matter, both versions of the model must give the same ground state and pressure.

Without diquark pairing (top panel), the pressure stays at its vacuum value (which is normalized to zero) at zero temperature until the baryon chemical potential reaches the lowest mass scale in the model (Silver-Blaze property \cite{Cohen:2003kd}), which is $3M_{\text{vac, light}}\simeq 1100\,$MeV (with the vacuum mass of the two light flavors $M_{\text{vac, light}}=367\,$MeV). As we also see in \cref{fig:dsdt23_T=0}, the higher temperature derivatives of the pressure are zero below this scale.
At $T=1\,$MeV and $T=50\,$ MeV in the model with diquark pairing (lower panel in \cref{fig:pressureT=0}), the phase transition between the 2SC and the CFL phase is first order (see \cref{fig:composition}, which manifests as a kink at $\mu_B\simeq 1340\,$MeV for $T=1\,$MeV and at $\mu_B\simeq 1400\,$MeV for $T=50$\,MeV, respectively, to a steeper slope of the pressure in the bottom panel of \cref{fig:pressureT=0}). Other than in unpaired quark matter (top panel), the $T=1$\,MeV and $T=50\,$MeV lines in the CFL phase (bottom panel) lie very close, indicating the suppression of thermal pressure due to the Cooper-pair quasiparticles.
At $T=75\,$MeV, the phase transitions are second order, such that the increase in slope is smooth, and at $T=100\,$MeV the ground state of the color-superconducting system is very close to the phase boundary to unpaired quark matter, and the lines in the top and bottom panels almost align.

We now calculate the coefficients needed for the Taylor expansion in \cref{eq:thermos}. As the entropy density is zero at $T=0$, we are left with calculating derivatives of the entropy density at constant baryon chemical potential and temperature,
\begin{equation}
    c_n=\frac{\partial^n P}{\partial T^n} \Big|_{T=0,\mu_B}.
\end{equation}
 The expressions for these derivatives in our model are derived in \cref{app:derivatives}. As a check, we tested the implementation of the coefficients using the formulas in \cref{app:derivatives} against the numerical derivatives using finite differences.
The first, second and third derivatives of the entropy with respect to temperature, taken at $T=1\,$MeV for easier numerical implementation, are shown in \cref{fig:dsdt23_T=0}. 
The first derivative is the specific heat of the system divided by temperature and as such effectively counts the degrees of freedom of the system. Its value jumps discontinuously from zero to a finite value at the point of the first order chiral symmetry breaking transition. 

For the unpaired matter (dark blue lines in Fig.\ \ref{fig:dsdt23_T=0}), $c_2$ increases continuously with chemical potential. The slope increases between $\mu_B \approx 1400\,$MeV and $1500\,$MeV due to the building up of a nonzero strange quark density (see right upper panel in \cref{fig:composition}). At even higher chemical potentials, the strange quark fraction $Y_s$ slowly approaches 1/3 and the derivative of the entropy increases with a smaller slope, which is, however, larger than the slope without strange quarks present.

In the model with diquark interactions (light blue line in Fig.\ \ref{fig:dsdt23_T=0} top), the phase transition between the chiral broken phase and the 2SC phase takes place already around $\mu_B\approx 1000\,$MeV (this can also be seen from \cref{fig:composition}). As we pointed out in \cref{sec:paired_expansion}, the contribution of paired modes is non-analytic in temperature around $T=0$, thus derivatives of the pressure with respect to temperature at low temperatures cannot capture the contributions by the paired modes. The derivative $c_2$ 
in the 2SC phase only captures the contribution of one unpaired color of up and down quarks (conventionally, blue quarks) and all three colors of strange quarks. Its value is therefore approximately 1/3 of the value of unpaired quark matter with a smaller slope which increases slightly around $\mu_B\approx1300\,$MeV, due to a tiny fraction of strange quarks $Y_s\lesssim 0.005$. At the first-order phase transition to the CFL phase, $c_2$ jumps to very small values close to zero as expected, as there are no unpaired modes which can store heat.

The $T^3$ coefficient, $c_3=\partial^2 s/ \partial T^2\vert_{T=0}$, is always  much smaller in magnitude than $c_2$, as shown in Fig.\ \ref{fig:dsdt23_T=0} in the middle plot. Unlike $c_2$, the coefficient $c_3$ may be positive or negative (as long as this term is overall smaller than the $\mathcal{O}(T^2)$ term). 
Indeed, this coefficient is exactly zero for a massless relativistic fermi gas and for a relativistic mean field model it was found to be negative but very small in magnitude in Ref.~\cite{Mroczek:2024sfp}. For our NJL model in the case without diquark pairing, it is non-zero at nearly all $\mu_B$ but is orders of magnitude smaller than the $c_2$ term. Around the onset of strange quarks it forms a double-peak structure with a positive and a negative value peak, where these peaks range from $\pm [10^3,10^4]$ MeV followed by an interval of negative values at $\mu_B\in[1400,1460]\,$MeV around the onset of strange quarks, before switching sign again after the crossover. A similar double peak with sign change at the onset of strange quarks is found for the model with diquark pairing (light blue line in Fig.\ \ref{fig:dsdt23_T=0} middle). In the CFL phase, the coefficient is zero, as expected. Thus the inclusion of the $T^3$ term in the expansion only changes our results close to the $ud\to uds$ crossover transition along the $\mu_B$ axis. These type of phase transitions, however, cannot be captured correctly by the Taylor expansion, so we do not include terms higher then order $T^2$ when comparing against the exact calculation of the model at nonzero temperature in the next subsection.

The dimensionless coefficient $c_4$, shown in the bottom plot of \cref{fig:dsdt23_T=0} shows a similar qualitative behavior as $c_3$, being positive in the absence of phase transitions, but showing peaks around the strange quark onset. Close to the strange quark onset, there are two additional sign changes, leading to an even more oscillatory behavior than $c_3$. Comparing the magnitude with the $T^2$ coefficient in the expansion, the ratio between the term proportional to $T^4$ and the term proportional to $T^2$ is $(c_4/c_2)\cdot T^2/12$. From \cref{fig:dsdt23_T=0} we can estimate that at high temperatures of order $T\sim 100\,$MeV, this ratio can be of the order of 10\% such that we expect an improvement of the $T$ expansion by going to order $T^4$ compared to order $T^2$ at the 10\% level, at least for unpaired quarks.

\subsection{Reconstructed finite \texorpdfstring{$T$}{T} EoS}\label{sec:reconstruction}

\begin{figure*}[htp]
    \centering
\centering
  \begin{minipage}[t]{0.49\textwidth}
    \centering
\includegraphics[width=\linewidth]{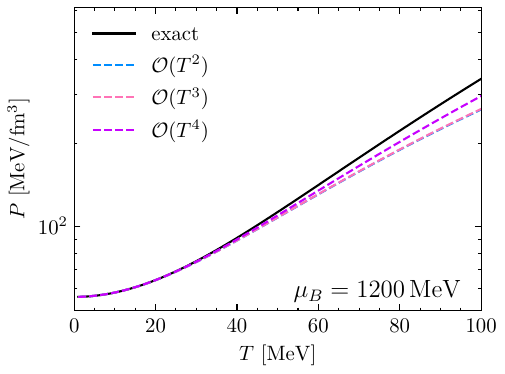}\\
\includegraphics[width=\linewidth]{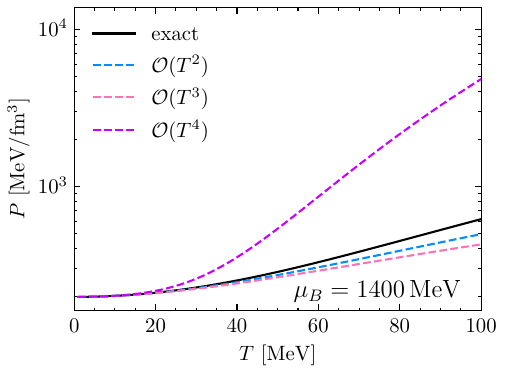}\\
\includegraphics[width=\linewidth]{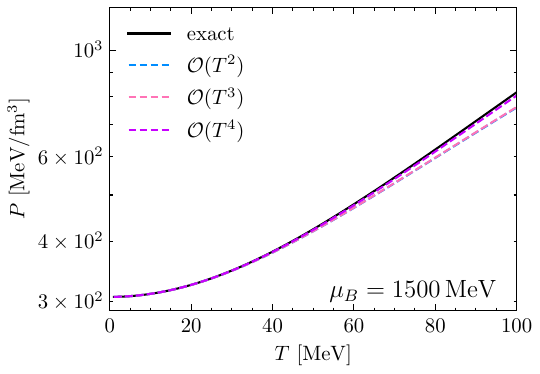}
  \end{minipage}\hfill
  \begin{minipage}[t]{0.49\textwidth}
    \centering
\includegraphics[width=\linewidth]{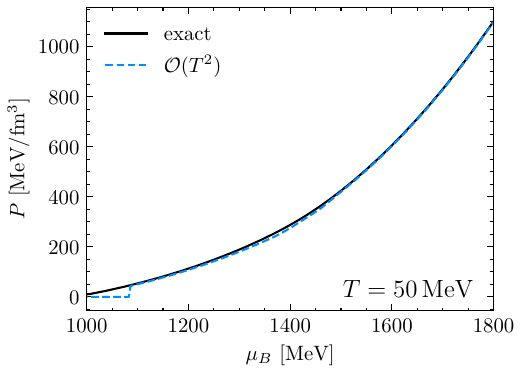}\\
\includegraphics[width=\linewidth]{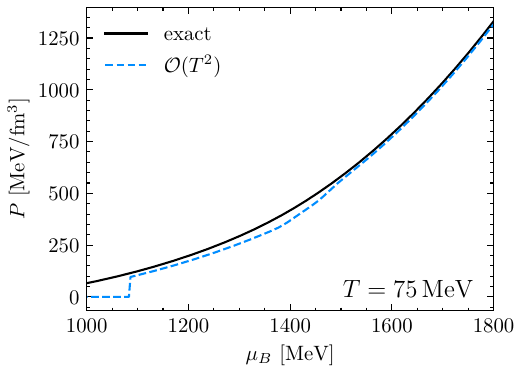}\\
\includegraphics[width=\linewidth]{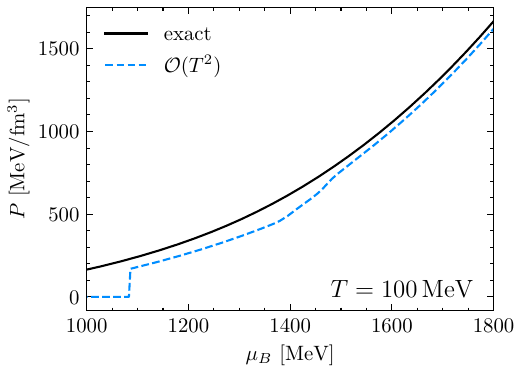}
  \end{minipage}
    \caption{From top to bottom: pressure versus temperature at fixed baryon chemical potentials $\mu_B=1200, 1400, 1500\,$MeV (left panels) and pressure versus baryon chemical potential at fixed temperatures $T=50,75,100\,$MeV (right panels) of the model without diquark interactions ($G_D=0$). The exact model calculation (black solid line) is compared with the Taylor expansion to order $T^2$ and orders $T^3$ and $T^4$ (only left panels). }
\label{fig:pressure_v_T_muB1200}
\end{figure*}

Now that we have calculated the entropy derivatives that contribute to the finite $T$ expansion, we can reconstruct the pressure at finite $T$ to test the range of validity of the expansion.

\subsubsection*{{\bf Finite $T$ expansion without diquark pairing}}
In \cref{fig:pressure_v_T_muB1200}, we plot the pressure vs temperature at three fixed values of $\mu_B=1200$\,MeV, $1400\,$MeV and $\mu_B=1500\,$MeV
in the left column and the pressure versus baryon chemical potential at three fixed temperatures $T=50$\,MeV, $75\,$MeV and 100\,MeV in the right columns for the model without diquark pairing ($G_D=0$). 
In all panels, the ``exact'' pressure calculated directly from the NJL model is shown in a solid black line. 
For the panels shown at fixed $\mu_B$ (left), we compare the exact calculation to the Taylor expansion \cref{eq:T_expansion_normal} up to order $\mathcal{O}((T/\mu_B)^2)$ (light blue blue dashed line), $\mathcal{O}((T/\mu_B)^3)$ (flamingo dashed line), $\mathcal{O}((T/\mu_B)^4)$ (purple dashed line) with the $c_2-c_4$ coefficients calculated in \cref{sec:coefficients}.  
For the right column, when comparing the exact solution to the Taylor series expansion at fixed temperatures, we only compare the Taylor expansion up to order $\mathcal{O}((T/\mu_B)^2)$.

Looking at the left columns plots in \cref{fig:pressure_v_T_muB1200}, we find that 
the expansion matches the exact calculation  well up to temperatures of $T\sim 50\,$MeV for $\mathcal{O}((T/\mu_B)^2)$ and more, depending on the chemical potential. At high temperatures, the pressure is under-estimated for $\mathcal{O}((T/\mu_B)^2)$ by the expansion with growing error. 
The expansion works best at large $\mu_B$ (bottom left plot in \cref{fig:pressure_v_T_muB1200}) compared to low $\mu_B$ (top left plot in \cref{fig:pressure_v_T_muB1200}).
The middle left panel in \cref{fig:pressure_v_T_muB1200} at $\mu_B=1400$ MeV, may at first be surprising in terms of its large deviations for the expansion at $\mathcal{O}((T/\mu_B)^4)$. Naively, including higher orders should improve fits, and we would normally expect higher $\mu_B$ to show improvements compared to the lower $\mu_B=1200$ MeV comparisons. 
However, it is important to compare this result to the $G_D=0$ phase diagram in \cref{fig:composition}, where we see that $\mu_B=1400$ MeV is just to the left of the crossover transition to strange quark matter where we start (at low $\mu_B$) with predominately $ud$ quarks and then strangeness appears such that we have  $uds$ quark matter. In particular, we see that fixing $\mu_B=1400$ MeV and increasing $T$ leads to the appearance of strange quarks at finite $T$ that did not exist at $T=0$.  
Thus, what we are seeing in the middle left panel in \cref{fig:pressure_v_T_muB1200} is that the expansion works well at low $T$ before strangeness becomes relevant but at high $T$ it breaks down because the $T=0$ limit for the $c_2-c_4$ coefficients did not encode information about the appearance of strangeness at finite $T$. We also note that the $\mathcal{O}((T/\mu_B)^2)$ truncation performs better at higher temperatures than the $\mathcal{O}((T/\mu_B)^4)$ result. This behavior points to the oscillatory nature of the expansion coefficients observed in \cref{fig:dsdt23_T=0} at $\mu_B\sim1400$MeV. As a consequence, including higher order coefficients does not systematically improve the approximation. In \cref{app:generic_stability} we discuss how a Taylor polynomial with negative higher order coefficients can in general lead to a thermodynamic instability at finite temperatures. In general, the expansion works worst at chemical potentials just below the appearance of strange quarks at $T=0$ (top middle plot in \cref{fig:pressure_v_T_muB1200}), where the strange quark mass drops and the contours of constant $Y_s$ are dense in the top right plot of \cref{fig:composition}.

Let us now turn to the right column in \cref{fig:pressure_v_T_muB1200} where we fix the $T$ and study the $\mu_B$ dependence of our expansion. Given that we already saw in the left column, that the order $\mathcal{O}((T/\mu_B)^2)$ is sufficient to describe most of the pressure (when the expansion is still in its regime of validity), we choose to only show the Taylor series expansion up to $\mathcal{O}((T/\mu_B)^2)$ for the fixed $T$ plots. 
Here we compare $T=50$ MeV (which we previously found reproduced the exact calculation well in the Taylor series) to higher values of $T=75$ MeV and $T=100$ MeV.

Below the chiral symmetry breaking phase transition at $\mu_B\simeq 1100\,$MeV, the expansion is zero due to the Silver-Blaze property mentioned in \cref{sec:coefficients}.
Thus, at this low $\mu_B$, the pressure also vanishes in the limit where $T$ goes to zero, i.e. $P(T=0)=0$, and the expansion cannot work.
This property implies that in an EoS where only the vacuum exists at $T=0$ but there is matter at finite $T$, then our Taylor series expansion breaks down because the $T=0$ limit does not provide any information (all the terms in the series are identically zero). 
However, this is also in the limit of low $\mu_B$ where we do not expect our Taylor series to be valid.

In the right column in \cref{fig:pressure_v_T_muB1200}, we find that, as expected, across all $\mu_B$ above the chiral symmetry breaking phase transition we reproduce the exact pressure well, especially for $T=50$ MeV (top), but even for $T=75$ MeV (middle) the result is very good. 
In fact, at high enough $\mu_B\gtrsim 1500$ MeV, the expansion at $\mathcal{O}((T/\mu_B)^2)$ is a good approximation of the exact solution at temperatures as high as 100 MeV. 
Given that our expansion can be thought of as a $T/\mu_B$ expansion (recall this is for symmetric matter where we can ignore other chemical potentials and $\mu_B$ is constant at each point in our expansion),  then we find that once the ratio of $T/\mu_B\gtrsim 0.05$ we begin to see small deviations from the $\mathcal{O}(T^2)$ results. 
Comparing both the left and right columns, we can anticipate that if we were to expand up to $\mathcal{O}((T/\mu_B)^4)$, we could provide a reasonable description even up to $T=100$\,MeV, with the only exception being around the $ud-uds$ matter crossover. 
Of course, in regimes where phase transitions or crossover exist and/or new degrees of freedom appear only at finite $T$ (e.g. strangeness) then the Taylor series also breaks down, as we have already pointed out. 

\begin{figure}[htp]
\centering
    \includegraphics[width=\linewidth]{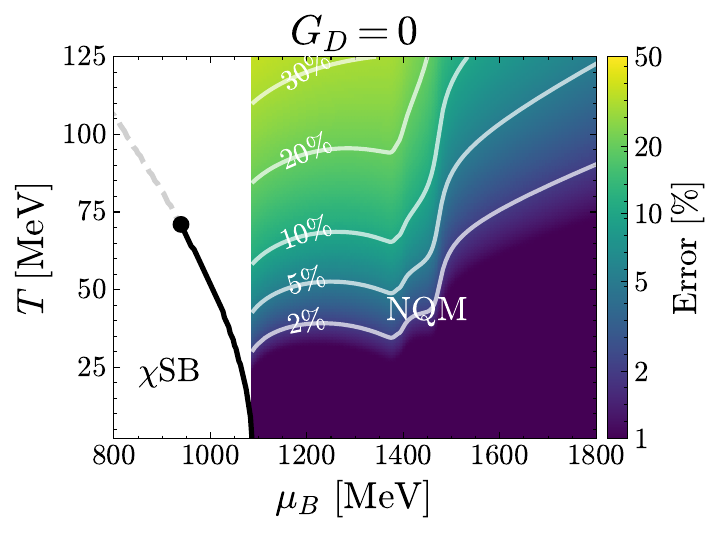}
     \caption{Absolute value of the relative error in the Taylor series expansion up to order $\mathcal{O}(T^2)$  compared to the exact NJL solution. The results are overlaid on tope of the phase diagram in the plane of baryon chemical potential and temperature of the model without diquark interactions.}
     \label{fig:heatmap_NOPAIR_error}
\end{figure}

In \cref{fig:heatmap_NOPAIR_error} we study the relative percentage error that we obtain in our expansion for $\mathcal{O}((T/\mu_B)^2)$. For $\mu_B<1100$\,MeV, we do not show the relative error because the pressure at $T=0$ vanishes due to the Silver Blaze effect. Therefore, any expansion from the $T=0$ point will only return vanishing pressure. 
However, at $\mu_B>1100$ MeV we found that the relative error percentage is low, and even up to $T=100$ MeV we only reach around $20\%$ relative error. 

Of course, at high $\mu_B\gtrsim 1500$ MeV, our expansion performs the best (as expected because $T/\mu_B\lesssim 0.06$), and even up to $T=100$ MeV our maximum relative error stays below $10\%$ (but often is significantly below that). 
At intermediate chemical potentials $1400\,\text{MeV}\lesssim \mu_B \lesssim 1500\,$MeV, we can visually see the effect of the crossover from light normal quark matter into strange normal quark matter, where the absolute relative error becomes worse over a very short range of $\mu_B$. However, even in the worst case the error only reaches up to about $20\%$ for $T=100$ MeV. Then, for the lowest $\mu_B$ we see again an increase in error which is to be expected because the $T/\mu_B\lesssim 0.09$ relation is larger than for the higher $\mu_B $ regime.   
If we limit ourselves to the regime where $T\le 50$ MeV, we find that the error is at most $\sim 5\%$. 

\begin{figure*}[htp]
\setlength{\tabcolsep}{0pt} 
\begin{tabular}{@{} m{0.46\textwidth} @{\hspace{0.04\textwidth}} m{0.46\textwidth} @{}}
    \begin{center}
      \includegraphics[height=0.245\textheight,keepaspectratio]{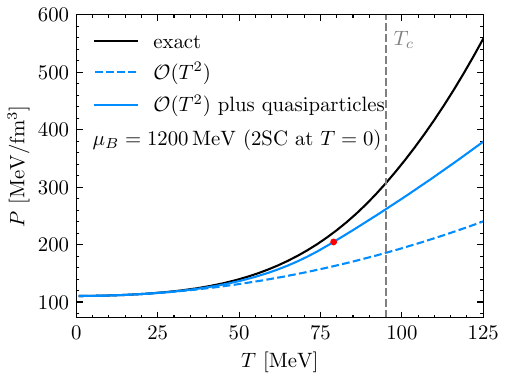}
    \end{center} &
    \begin{center}
      \includegraphics[height=0.24\textheight,keepaspectratio]{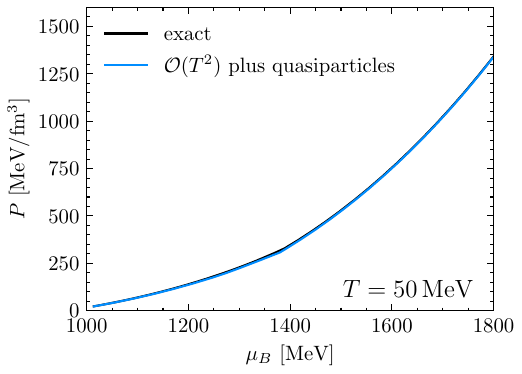}
    \end{center} \\[1ex]

    \begin{center}
      \includegraphics[height=0.245\textheight,keepaspectratio]{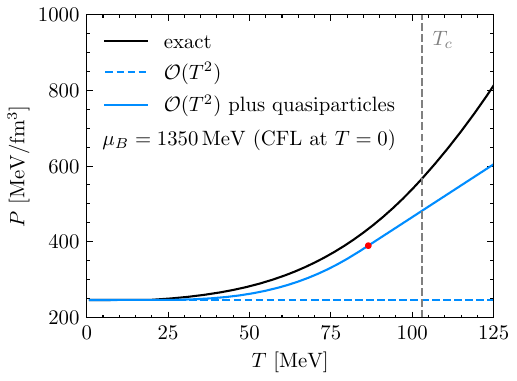}
    \end{center} &
    \begin{center}
      \includegraphics[height=0.24\textheight,keepaspectratio]{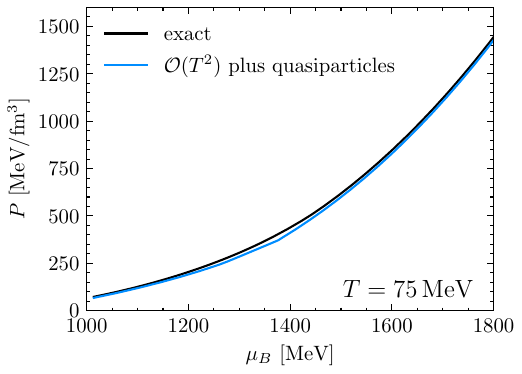}
    \end{center} \\[1ex]

    \begin{center}
      \includegraphics[height=0.245\textheight,keepaspectratio]{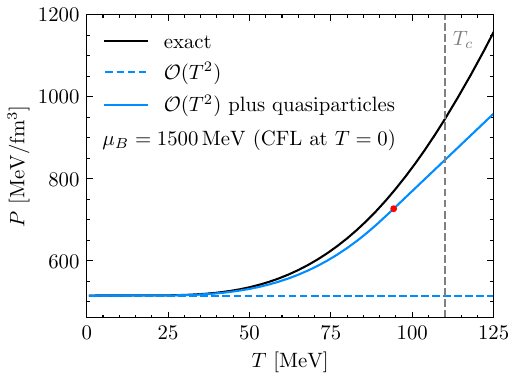}
    \end{center} &
    \begin{center}
      \includegraphics[height=0.24\textheight,keepaspectratio]{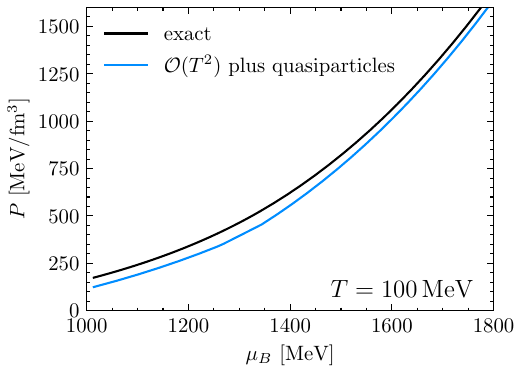}
    \end{center}
\end{tabular}

    \caption{From top to bottom: pressure versus temperature at fixed baryon chemical potentials $\mu_B=1200, 1350, 1500\,$MeV (left panels) and pressure versus baryon chemical potential at fixed temperatures $T=50,75,100\,$MeV (right panels) of the model with diquark interactions ($G_D=G_S$). Left: the exact model calculation (black solid line) is compared with the Taylor expansion to order $T^2$ (blue dashed) and the full expansion including the term for quasiparticles \cref{eq:general_pairings} (blue solid). For temperatures $T>T^\ast$, with $T^\ast$ indicated by the red dot, the constant entropy density extrapolation \cref{eq:extrapolation} is applied. Right: the exact model calculation (black solid line) is compared with the full expansion of \cref{eq:extrapolation} (blue solid).
    }
\label{fig:pressure_validation}
\end{figure*}

\subsubsection*{{\bf Finite $T$ expansion with diquark pairing}}
In \cref{fig:pressure_validation}, we show a similar plot as \cref{fig:pressure_v_T_muB1200} but for the temperature and chemical potential dependence of the pressure for the model including diquark pairing ($G_D=G_S$). 
Once again, the left column fixes the $\mu_B$ at specific values and compares the $T$ dependence of the exact vs expanded values whereas the right column fixes the $T$ and focuses on the $\mu_B$ dependence. 
Here we do not choose the exact same fixed $\mu_B$ slices as in \cref{fig:pressure_v_T_muB1200}, because we want to be careful about selecting certain points of interest in the phase diagram with diquark pairing, shown in \cref{fig:composition}.
Thus, the chemical potentials for the left plots are chosen where the zero temperature phase is 2SC ($\mu_B=1200\,$MeV, top), CFL ($\mu_B=1500\,$MeV, bottom), and in the CFL phase at a chemical potential just above the 2SC-CFL phase transition ($\mu_B=1350\,$MeV, middle). 
However, the $T$ slices are the same as the ones previously shown in \cref{fig:pressure_v_T_muB1200}. 

Since we already found that $\mathcal{O}((T/\mu_B)^2)$ is reasonably sufficient to describe the pressure up to $T\sim 50$ MeV in  \cref{fig:pressure_v_T_muB1200}, we only make comparisons up to $\mathcal{O}((T/\mu_B)^2)$ in \cref{fig:pressure_validation}. 
However, because we have color-superconducting phases that appear we compare the exact calculation with the Taylor expansion up to order $\mathcal{O}(T^2)$ (dashed) and the full approximation for color-superconducting phases \cref{eq:general_pairings} (solid) which adds a term for the quasiparticle modes to the Taylor expansion. The temperature dependence of the diquark gap was parameterized using the BCS expression \cref{eq:Delta_T_approx}.

The Taylor expansion cannot capture the contributions of paired modes, as they are non-analytic in temperature. Thus, in the CFL phase, all Taylor coefficients vanish and the Taylor expansion does not work. In the 2SC phase, the Taylor expansion captures at least the contribution of the the blue unpaired up and down quarks and the few strange quarks which are present at $T=0$ which gives a useful result up to $T\simeq50\,$MeV (see the dashed line in the top left plot in \cref{fig:pressure_validation}) with an increasing under-estimation of the pressure at higher temperatures.
Away from the 2SC-CFL transition (middle and bottom left plots), the expansion including the contribution of the quasiparticles follows the pressure increase in temperature compared to the exact pressure well. 
For \cref{eq:Delta_T_approx}, the critical temperature $T_c$ is estimated from the zero temperature gap $\Delta(T=0)$ using the relation \cref{eq:Tc_approx}. The exact critical temperature is of the order of 10\,MeV above this value and is shown as a vertical dashed gray line in the left panels of \cref{fig:pressure_validation}.

We now focus on the right column in \cref{fig:pressure_validation} where we fix the same temperatures as done previously for the unpaired quarks, and study the $\mu_B$ validity of our expansion. In the fixed $T$ plots, we do not show results without the quasiparticle term, as its inclusion is necessary in color-superconducting phases. At chemical potentials above $\mu_B\gtrsim 1340\,$\text{MeV} the 2SC-CFL phase boundary bends to increasing $\mu_B$ with increasing $T$ (cf. \cref{fig:composition}, right plots). The reconstruction at $\mu_B$ directly after this transition changes from the formula for 2SC quasiparticles, to the formula for CFL quasiparticles (cf. \cref{tab:T_exp}). Applied naively, this can lead to non-monotonic behavior of the pressure as a function of $\mu_B$ at certain temperatures, which would imply negative $n_B$ and is unstable. This artifact in our reconstruction is only due to the missing information about the bending of the 2SC-CFL phase boundary at nonzero temperature and we solve it by a simple linear interpolation, see \cref{app:smooth}. Our results shown in \cref{fig:pressure_validation} and subsequent plots include this interpolation.

Overall, we find that this procedure works well for almost all chemical potentials (see also the right column plots in \cref{fig:pressure_validation}) until $T=75\,$MeV (both the top right and top middle plots). At $T=100\,$MeV (bottom right plot) the reconstructed pressure has been calculated with the constant entropy density extrapolation, but still follows the exact value accurately for $\mu_B\gtrsim1500\,$MeV.

The worst performance of the approximation is between $\mu_B=1340\,$MeV and $\mu_B=1400\,$MeV where the 2SC-CFL transition is first order (see the middle left plot and all plots in that $\mu_B$ range in the right panels). At $\mu_B=1350\,$MeV (middle left), the first-order 2SC-CFL transition happens already at $T\approx20$\,MeV. Our formula \cref{eq:general_pairings} misses this information but still performs adequately with only a small underestimation of the pressure at the $10\%$ level until $T=80\,$MeV. Only at around $T=100\,$MeV, close to the critical temperature of the 2SC phase, the deviations become large, see the bottom, right plot in \cref{fig:pressure_validation}.

\begin{figure}

    \includegraphics[width=\linewidth]{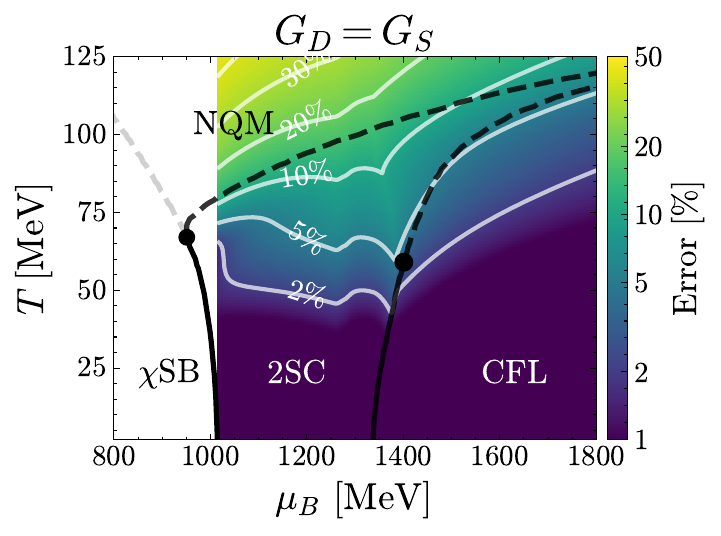}
\caption{
      Absolute value of the relative error in the reconstructed pressure using the Taylor series expansion to order $T^2$ plus the expression for the quasiparticles \cref{eq:extrapolation} compared to the numerical mean-field solution of the NJL model with diquark pairing ($G_D=G_S$). The results are overlaid on top of the phase diagram in the plane of baryon chemical potential and temperature.}
     \label{fig:heatmap_CSC_error}
    \end{figure}

The absolute value of the relative error of the expansion for paired quarks including the Taylor expansion to $\mathcal{O}(T^2)$ and the term for the quasiparticles in the phase diagram is shown in \cref{fig:heatmap_CSC_error}. Compared to \cref{fig:heatmap_NOPAIR_error} the phase diagram of the model with diquark pairing is more complex: there is a second critical point at $T\simeq 60$\,MeV and second order phase transitions corresponding to 2SC unpairing and CFL unpairing across all $\mu_B$. 
Despite this, we find quite similar results for the magnitude of the relative error: we once again find that up to $T\sim 50$ MeV (relevant for neutron star mergers) our error is at most at a few percent level. At high $T\sim 100$ MeV, the error is more than $20\%$ in normal quark matter, but decreases with increasing $\mu_B$ to less than $10\%$ in the 2SC and CFL phase. As in the case of unpaired matter, the thermal pressure of three-flavor color-superconducting matter is better approximated than for two-flavor color-superconducting matter.  Of course, the phase structure itself plays a crucial role in our relative error, as we discuss in detail next.

 The line of $5\%$ error closely follows the second order 2SC-CFL phase boundary. Where the 2SC-CFL transition is first order, the lines of constant error jumps (see e.g. the 2$\%$ error line). A kink of the error contours around $\mu_B\simeq1350\,$MeV is also visible at higher temperatures (see the $10\%$ error contour). This is an artifact of the interpolation we used to connect the 2SC and CFL pressure expansion, see \cref{app:smooth}. Overall the relative error stays at the few-percent level (typically \(\lesssim 5\text{–}10\%\)) across most of the 2SC and CFL domains, consistent with \cref{fig:pressure_validation}. 
\subsection{Thermal index}\label{sec:thermal_index}

\begin{figure*}
    \centering
    \begin{tabular}{cc}
       \includegraphics[width=0.47\linewidth]{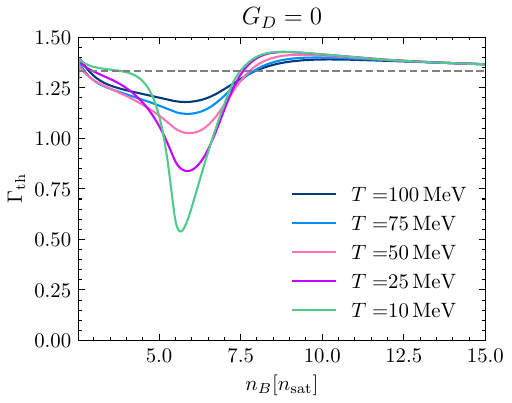}  &  \includegraphics[width=0.5\linewidth]{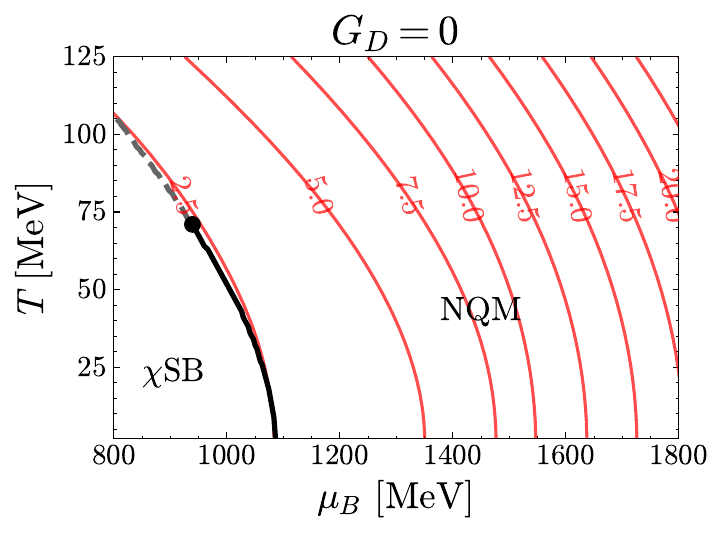}  \\
    \end{tabular}
    \caption{The thermal index vs net baryon density (left) and the corresponding phase diagram  in $T,\mu_B$ with the isodensity contours drawn that correspond to points on the $n_B$ axis in the thermal index plot. The thermal index results are shown for different choices of fixed temperature.  The results are shown for unpaired quarks only in the NQM phases where there is a crossover from $ud$ to $uds$ quark matter.}
    \label{fig:thermalindex_GD0}
\end{figure*}

A common method to approximate thermal effects is by calculating the thermal index \cite{Janka:1993,Bauswein:2010dn}:
\begin{align}
    \Gamma_{\text{th}}=1+\frac{P_{\text{th}}}{\varepsilon_{\text{th}}}
\end{align}
where $P_{\text{th}}(T)=P(T)-P(T=0)$ and $\varepsilon_{\text{th}}(T)=\varepsilon(T)-\varepsilon(T=0)$ are the thermal pressure and thermal energy density, respectively.
In a numerical relativity simulation, the thermal energy density is evolved together with the baryon number density. Assuming a temperature-independent thermal index, the thermal pressure can then be approximated from:
\begin{align}
    P_{\text{th}}=(\Gamma_{\text{th}}-1) \varepsilon_{\text{th}},
\end{align}
instead of looking up the value of the pressure in a temperature-tabulated equation of state. The values for a non-relativistic and a relativistic ideal gas are $\Gamma_{\text{th}}=5/3$ and $\Gamma_{\text{th}}=4/3$, respectively. For hadronic EoSs without hyperons $\Gamma_{\text{th}}=1.75$ is a reasonable approximation \cite{Bauswein:2010dn}, while for unpaired quark matter typically the value of a massless relativistic gas $\Gamma_{\text{th}}=4/3$ is used, e.g. Ref.~\cite{Blacker:2023afl}.
However, the thermal index is rather non-trivial for hyperons, see Ref.~\cite{Blacker:2023opp,Kochankovski:2025lqc}, where the thermal index may have a stronger density dependence than for nucleons \cite{Constantinou:2015mna}.

\begin{figure*}
    \centering
    \begin{tabular}{cc}
       \includegraphics[width=0.47\linewidth]{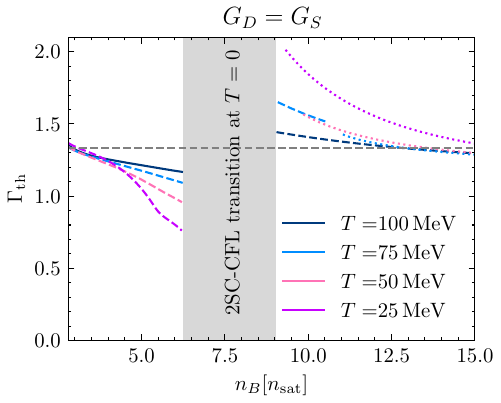}  &  \includegraphics[width=0.5\linewidth]{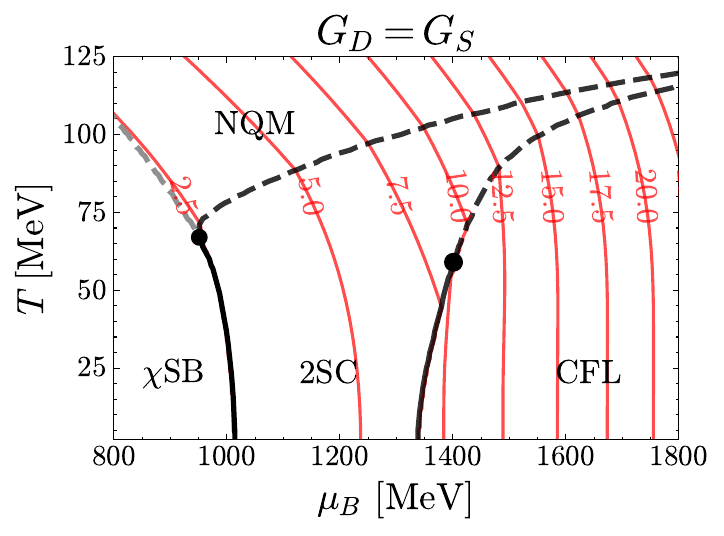}  \\
    \end{tabular}
    \caption{The thermal index vs net baryon density (left) and the corresponding phase diagram  in $T,\mu_B$ with the isodensity contours drawn that correspond to points on the $n_B$ axis in the thermal index figure. The thermal index results are shown for different choices of fixed temperature. Because of the non-trivial relationship between $n_B(\mu_B)$, lines of $T=\text{const.}$ cross multiple different phases of matter and are indicated by different line styles in the left plot. Depending on $n_B$ and $T$, matter is either in the NQM phase (solid lines), the 2SC phase (dashed lines) or the CFL phases (dotted lines). The discontinuity of the curves in the left plot is due to the first-order 2SC-CFL transition at low $T$.
    }
    \label{fig:thermalindex_pair}
\end{figure*}

In \cref{fig:thermalindex_GD0}, we plot the thermal index for unpaired quarks (left) along with their corresponding phase diagram (right) where isodensity lines (corresponding to points on the $n_B$-axis in the thermal index plot) are highlighted in red \footnote{The densities $n_B\gtrsim 8\,n_{\text{sat}}$ at which $uds$ quark matter in \cref{fig:thermalindex_GD0} and the CFL phase in \cref{fig:thermalindex_pair} appear, respectively, might seem not reachable in binary neutron star mergers. Note, however, that including a repulsive vector interaction shifts the onset of strange matter and the CFL transition to lower densities $n_B\sim 3-5\,n_{\text{sat}}$, see e.g. the phase diagram in Ref.~\cite{Alford:2025jtm}. In this work, we focus on presenting the finite $T$ framework and do not include repulsive vector interactions for simplicity.}. The gray dashed line (left) indicates the ideal, relativistic massless gas value of $\Gamma_{\text{th}}^{\text{ideal}}=4/3$.  The thermal index is plotted along lines of $T=\text{const.}$ for a variety of temperatures of interest. 
We find that at high $n_B$ all $T$ slices converge to a thermal index that is consistent with an ideal, massless gas. This has to be the case, as our RG-consistent setup fulfills the Stefan-Boltzmann limit for $n_B\to\infty$.

At low $n_B$, our minimum density corresponds to a range within $ud$-dominated NQM  where the strange quark has a quite large constituent mass (such that $Y_s$ is small). In that regime, our system is not conformal, but we do find that our different temperatures all converge to a similar value of the thermal index of $\Gamma_{\text{th}}\sim 1.4$ at very low $n_B$.  

As we vary $n_B$ along lines of fixed $T$, we see that there is a strong $n_B$ dependence. Essentially, we find a large dip in $\Gamma_{\text{th}}$ around $n_B\sim 5-7\,n_{\text{sat}}$, which corresponds to the crossover regime where strange quarks rapidly appear within NQM. The dip is strongly $T$ dependent: at $T=10\,$MeV it reaches nearly to $\Gamma_{\text{th}}=0.5$ while the dip at high $T$ is  more subtle. 
Such a decrease of the thermal index at intermediate densities due to strange degrees of freedom has also been seen for equations of state with hyperons \cite{Blacker:2023opp,Kochankovski:2025lqc}. 

While the phase structure is significantly less complicated for NQM matter (unpaired quarks) than in the case with paired quarks, we can clearly see in \cref{fig:thermalindex_GD0} that the thermal index has a strong $T,n_B$ dependence. Therefore, a single value of $\Gamma_{\rm th}$ cannot approximate the pressure at finite $T$. 
Thus, even in the simpler case of unpaired quarks, a thermal index would fail to describe the model. 

We now turn to the more complicated case of  color-superconducting quark matter  in \cref{fig:thermalindex_pair}. There, we plot the thermal index at different $T$ slices for paired quarks (left) along with their corresponding phase diagram (right).   
Interestingly, in the low $n_B$ limit, the thermal index seems to converge to a value slightly above that of the ideal, massless relativistic gas (as we previously saw in \cref{fig:thermalindex_GD0} as well), despite the fact that two distinct phases appear at different temperatures (NQM at high $T$ and 2SC at low $T$). Thus it appears that in $ud$ quark matter at low densities the thermal index does not strongly depend on the phase of matter.
However, at higher $n_B$, we see a much stronger dependence on $T$ already at $n_B\sim 5n_{\text{sat}}$. The higher $T$ (in the NQM phase) shows a smaller $n_B$ dependence and is closest to the ideal limit whereas the low $T$ regime that is in the 2SC phases is strongly $T,n_B$ dependent and decreases down to low values even below $1$.

Between $\mu_B\simeq 1340\,$MeV and $\mu_B\simeq 1400\,$MeV, corresponding to densities around $\sim[6,10]\,n_{\text{sat}}$, CFL matter melts to 2SC via a first-order phase transition at a fixed $n_B$ with increasing temperature.
All lines in \cref{fig:thermalindex_pair} abruptly end around $n_B\sim 6.2\,n_{\text{sat}}$ because of this first-order phase transition that also occurs at $T=0$ such that there is a jump in the baryon density from $n_B\sim 6.2\,n_{\text{sat}}$ until $n_B\sim 9n_{\text{sat}}$ (indicated as a gray box in the left of \cref{fig:thermalindex_pair}). Because of the jump in $n_B$ at $T=0$ we do not have data to calculate the thermal index for that $n_B$ range. 

At densities above the transition, the lines of constant temperature in \cref{fig:thermalindex_pair}, left, lie in the CFL phase for low temperatures (dotted lines) or (partially) in the 2SC phase (dashed lines) for higher temperatures. For both cases, we find significant deviations from the ideal, massless relativistic gas thermal index, and at densities around $n_B\sim 9n_{\text{sat}}$ our thermal index is large $\Gamma_{\text{th}}\sim [1.5,2]$. 
The regime where $\Gamma_{\text{th}}$ is calculated entirely within the CFL leads to the largest thermal index (occurs for the lower $T$ slices), while the thermal index is smaller and flatter for the $\Gamma_{\text{th}}$ calculated at high $T$ where the phase boundary from CFL to 2SC is crossed. 
At lower temperatures of $T\simeq10\,$MeV the calculation of $\Gamma_{\text{th}}$  becomes unstable in the CFL phase. This can be understood again with the specific thermal behavior of the CFL phase at low temperatures, see \cref{fig:pressure_validation} middle and bottom left plots: the pressure and the energy density (not shown) are exponentially suppressed in temperature by the diquark gaps, staying at their constant $T=0$ values until the temperature is increased above a few tens of MeV. 
At these low temperatures, $\Gamma_{\text{th}}$ is numerically unstable, as both $P_{\text{th}}$ and $\varepsilon_{\text{th}}$ are zero.
At high $n_B\sim 15 n_{\text{sat}}$, we find that $\Gamma_{\text{th}}$ at different temperatures start to approach the same value, but with significantly less convergence than what we saw previously for the unpaired case. In fact, there is still a clear difference between the $T=25$ MeV line compared to higher temperatures.

These results highlight that a thermal index cannot easily capture the finite $T$ behavior of color-flavor locked quark matter. 
Moreover, the thermal index cannot even be calculated across a significant range of $n_B$ when crossing a first-order phase transition at $T=0$, and even when it is possible to calculate, we find significant differences in $\Gamma_{\text{th}}$ at different slices of $T$, especially for the lower temperatures. 
In contrast, our framework which combines the finite T expansion with the information from the gapped phases provides a significant improvement that is also more flexible to describe the complicated phase structure of dense quark matter.  
\section{Conclusions and Outlook}\label{sec:conclusions}

In this work we developed a new finite $T$ framework that can accurately reproduce the EoS of a three-flavor NJL model in the mean-field approximation. We build on the work Ref.~\cite{Mroczek:2024sfp} that was based on a Taylor expansion of the pressure at fixed chemical potentials around $T=0$. For color-superconducting phases, we extended this approach to include the thermal pressure of quasiparticles, which is non-analytic at $T\to0$, while enforcing thermodynamic stability. The extension is given by a simple analytic formula \cref{eq:extrapolation} which needs as additional input only the $ud$ pairing diquark gap at vanishing temperature $\Delta_3(T=0)$ and the chemical potentials $\mu_1$ and $\mu_2$ of the paired quark species.
We calculated the required coefficients for the finite $T$ framework and tested the results against the NJL mean-field calculations both with and without diquark pairing. In both cases, the phase diagrams show non-trivial features due to the temperature- and density dependence of the strange-quark fraction.

In the case without diquark pairing, the reconstructed finite $T$ EoS up to $\mathcal{O}(T^2)$ stays accurate to within $\simeq5\%$ for $T<50\,$MeV and to within $\simeq20\%$ for $T<100\,$MeV, with much lower errors with increasing densities.
In the case with color-superconductivity, where we use the Taylor expansion to $\mathcal{O}(T^2)$ together with the term for the thermal pressure of quasiparticles \cref{eq:extrapolation}, we find similar errors, even though the phase structure is more complicated than in the case without diquark pairing.
In neutron star merger simulations, temperatures mostly reach only up to about $T\approx 50$\,MeV and the maximum relative error in that regime remains mostly below $\sim5\%$.

There are two main challenges that we faced: the specific thermal behavior of quasiparticles in color-superconductivity and the appearance of strange degrees of freedom only at nonzero but not at zero $T$ for a given chemical potential. The Taylor series expansion alone fails to capture any finite $T$ effects of the quasiparticles, because they result in vanishing expansion coefficients i.e. $c_n=0$. However, with our analytic formula for the quasiparticle pressure \cref{eq:extrapolation}, we can describe the finite $T$ behavior of the color-superconducting phases sufficiently well. The second challenge is the appearance of strange quarks, especially when they appear only at finite $T$ for a fixed $\mu_B$, because the $T=0$ limit does not correctly capture their behavior. We note that this second issue may be improved on by taking strangeness as a separate degree of freedom in our expansion, but we leave that for future work. These two points make clear that a simple addition of finite $T$ effects through a constant thermal index would utterly fail to describe the EoS of the model at finite $T$. 

This paper now opens up the door to a number of interesting possibilities when it comes to understanding the applicability of EoS expansions at the low $T$, high $\mu_B$ end of the QCD phase diagram. In this work, we only compared the framework to the NJL mean-field results at isospin-symmetric matter. However, our work could be extended to quark matter with nonzero isospin asymmetry \cite{Danhoni:2025qpn} to more closely model beta-equilibrated quark matter in neutron stars. The beta-equilibrated neutral ground state for the CFL phase has $\mu_Q=0$ at zero temperature \cite{Rajagopal:2000ff} such that our expansion for the CFL phase stays the same. For 2SC quark matter, the formula \cref{eq:result_mismatch} for mismatched chemical potentials for up and down quarks has to be used in this case. Additionally, the expansion could be tested against other calculations of the phase diagram of dense quark matter, e.g. based on the renormalizable quark-meson diquark (QMD) model \cite{Braun:2018svj,Andersen:2024qus,Andersen:2025ezj,Andersen:2025uzh,Gholami:2025afm}. In the mean-field approximation, the spectrum of excitations in the QMD model is of a similar form as in the NJL model, thus we expect the expansion to work there as well. Furthermore, it would be interesting to test the expansion against QCD-based approaches which include quantum fluctuations beyond the mean-field approximation \cite{Braun:2020bhy,Braun:2022olp,Topfel:2024iop,Geissel:2024nmx,Stoll:2025jor,Geissel:2025vnp}.
Finally, we comment that for  unpaired quark matter the expansion can be compared to the recently proposed expansion based on Fermi liquid theory \cite{Zhu:2025vmz}.

Future work may also explore a more complex NJL model including more interaction channels, see e.g. Ref.~\cite{Yang:2025ach}.
It is necessary to include a repulsive vector interaction (e.g. Ref.~\cite{Klahn:2006iw}) in order to make the EoS stiff enough to support $2\,M_{\odot}$ neutron stars. This has been done already for the RG-consistent NJL model \cite{Gholami:2024ety,Christian:2025dhe} but was left out here for simplicity.

We also neglected pressure contributions from particles other than quarks and BCS quasiparticles. The CFL phase breaks chiral symmetry and baryon number conservation \cite{Alford:1998mk}, with the former leading to a pseudoscalar octet with inverted mass ordering \cite{Son:1999cm} and the latter to a massless bosonic superfluid mode \cite{Fukushima:2005gt}. While the thermal pressure of these particles could be just added as a boson gas, it is further predicted, that kaon condensation is favored in certain areas of the phase diagram of dense quark matter, especially for the CFL phase \cite{Schafer:2000ew,Warringa:2006dk,Basler:2009vk}. This would change the quasiparticle spectrum and thus the term we include for the thermal pressure of the quasiparticles in our approach.
Thus, different parameter sets and/or extensions of our NJL model may change the order of magnitude and/or behavior of our extracted $c_n$ coefficients that would be interesting to study.

In summary, the finite $T$ framework used here provides an easy-to-use way to include thermal effects into the cold equation of state of quark matter. Assuming a calculated or parameterized normal conducting or color-superconducting cold quark matter EoS, the thermal part of the EoS can be simply added, provided that the pressure, the temperature derivative of the entropy density $c_2$ and the diquark gap and chemical potentials of paired quarks at $T=0$ is given or assumed. We expect our work to be particularly useful for numerical relativity simulations of core-collapse supernova and neutron star mergers in which EoSs with different interactions and thermal effects are explored to find signatures of these different EoS properties in astrophysical observables. Highlighting the characteristic suppression of the thermal pressure due to the diquark gaps, associated also with characteristic transport properties \cite{Alford:2006gy,Wang:2010ydb,Berdermann:2016mwt,Alford:2024tyj,Alford:2025tbp,Carter:2000xf,Reddy:2002xc,Kundu:2004mz,Alford:2025jtm}, we propose that numerical relativity simulations of systems with dense color-superconducting quark matter could provide important insights.

\section{Acknowledgments}
We thank Andreas Bauswein, Liam Brodie, Michael Buballa, Alexander Haber, Lennart Kurth and Mateus Reinke Pelicer for helpful discussions.
We acknowledge support from the support from the US-DOE Nuclear Science Grant No. DE-SC0023861 and the National Science Foundation under the MUSES collaboration
OAC-2103680. H.G. and M.H. acknowledge support from the Deutsche Forschungsgemeinschaft (DFG, German Research Foundation) 
through the CRC-TR211 `Strong-interaction matter under extreme conditions' project number 315477589 – TRR 211. M.H. is supported by the GSI F\&E. D.M. is supported by
the National Science Foundation Graduate Research Fellowship Program under Grant No. DGE-1746047 and
the Illinois Center for Advanced Studies of the Universe Graduate Fellowship.

\onecolumngrid

\appendix

\section{Expressions for the coefficients of the finite temperature expansion}

\subsection{Pressure contribution from quasiparticles}\label{app:paired_contr}
We approximate the thermal pressure of quasiparticles analogously to a calculation in Ref.~\cite{Fukushima:2004zq} and a calculation for the specific heat in Ref.~\cite{Schmitt:2010pn}.
When the chemical potentials of quarks differ,
\(\mu_1 \neq \mu_2\),
we define the average and mismatch
\begin{align}
    \bar{\mu} &= \frac{\mu_1 + \mu_2}{2}, &
    \delta\mu &= \mu_1 - \mu_2.
\end{align}
The quasiparticles then have two dispersion branches,
\begin{align}
    \epsilon_{\pm}(p) = \sqrt{(p - \bar{\mu})^2 + \Delta^2} \pm \frac{\delta\mu}{2},
\end{align}
corresponding to the ``breached'' and ``anti-breached'' modes. The explicit temperature-dependent pressure contribution from these modes is
\begin{align}
    P^\text{paired}_{\text{th}}
    &= \frac{T}{2\pi^2}\sum_{\sigma=\pm}
    \int_0^\infty p^2 dp\,
    \ln\!\left[1+\exp\!\left(-\frac{\sqrt{(p-\bar{\mu})^2+\Delta^2}+\sigma\,\delta\mu/2}{T}\right)\right].
\end{align}
Focusing on momenta close to the average Fermi surface \(p\simeq\bar{\mu}\) and setting \(x=(p-\bar{\mu})/T\),
\begin{align}
    P^\text{paired}_{\text{th}}
    &\approx \frac{\bar{\mu}^2T^2}{2\pi^2}
    \sum_{\sigma=\pm}\int_{-\frac{\bar{\mu}}{T}}^{\infty} dx\,
    \ln\!\left[1+\exp\!\left(-\sqrt{x^2+(\Delta/T)^2}-\sigma\delta\mu/(2T)\right)\right].
\end{align}
We approximate the lower integration boundary $-\frac{\bar{\mu}}{T}\approx -\infty$ and use that the integrand is symmetric in $x$:
\begin{align}
    P^\text{paired}_{\text{th}}
    &\approx \frac{\bar{\mu}^2T^2}{\pi^2}
    \sum_{\sigma=\pm}\int_{0}^{\infty} dx\,
    \ln\!\left[1+\exp\!\left(-\sqrt{x^2+(\Delta/T)^2}-\sigma\delta\mu/(2T)\right)\right].\label{eq:pairedpressureint}
\end{align}

For \(T\ll \Delta\), we may approximate \(\sqrt{x^2+(\Delta/T)^2}\simeq \Delta/T + x^2T/(2\Delta)\) and
\(\ln(1+e^{-y})\simeq e^{-y}\left(1-\tfrac{1}{2}e^{-y}\right)\) for large $y$, provided that
\begin{align}
    \Delta\ge\frac{\delta\mu}{2},
\end{align}
i.e., the system is not in a gapless color-superconducting phase \cite{Shovkovy:2003uu,Alford:2003fq}. This approximation gives
\begin{align}
P^\text{paired}_{\text{th}}\approx &
    \frac{\bar{\mu}^2T^2}{\pi^2}e^{-\Delta/ T}\sum_{\sigma=\pm}\exp{\left(-\frac{\sigma\,\delta\mu/2}{T}\right)}
    \int_0^\infty  dx \left[\exp{\left(-\frac{x^2}{2(\Delta / T)}\right)} - \frac{1}{2}e^{-\Delta/T}\exp{\left(-\frac{\sigma\,\delta\mu/2}{T}\right)}\exp{\left(-\frac{x^2}{\Delta / T}\right)}\right],
\end{align}
which can be solved analytically, yielding
\begin{align}\label{eq:result_mismatch}
    P^\text{paired}_{\text{th}}
    \approx
    \frac{\bar{\mu}^2 T^2}{\pi^{3/2}}
    \sqrt{\frac{\Delta}{T}}\,
    e^{-\Delta/T}
    \left[\sqrt{2}\cosh\!\left(\frac{\delta\mu}{2T}\right)
    - \frac{1}{2}e^{-\Delta/T}\cosh\!\left(\frac{\delta\mu}{T}\right)\right].
\end{align}
In the limit \(\delta\mu \to 0\), which is true for isopsin-symmetric matter at $T=0$, this result reduces to
\begin{align}\label{eq:paired_app}
P^\text{paired}_{\text{th}}\approx &
    \frac{\bar{\mu}^2}{\pi^{3/2}}T^2
    \sqrt{\frac{\Delta}{T}}\,\, e^{-{\Delta}/{T}}\,\left(\sqrt{2}-\frac{1}{2}e^{-{\Delta}/{T}}\right).
\end{align}
Note that a related expression for the thermal pressure of quasiparticles at low temperatures was derived in Ref.~\cite{Fukushima:2004zq}.

\section{Test of the parameterized temperature dependence of the diquark gap}\label{app:Delta_t}

\begin{figure*}[h]
    \begin{tabular}{cc}
\includegraphics[width=0.5\linewidth]{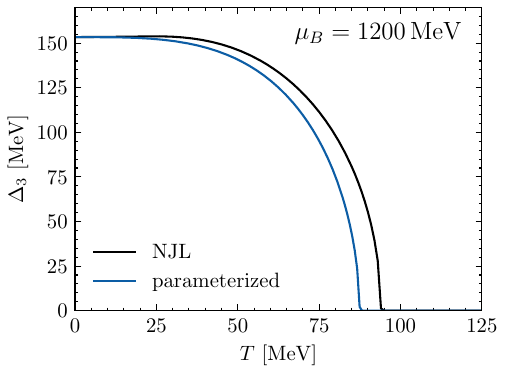}    &  \includegraphics[width=0.5\linewidth]{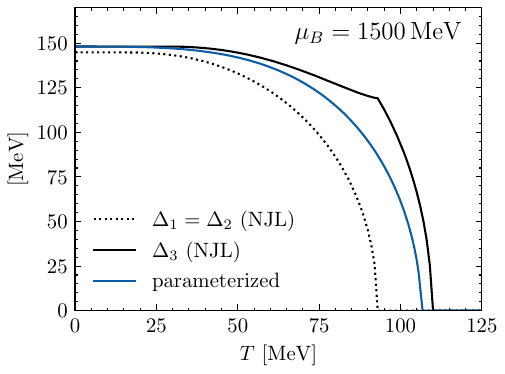}
    \end{tabular}
    \caption{Parameterized temperature dependence of the diquark gap (blue) compared to the calculation in the NJL model with $G_D=G_S$ (black) at $\mu_B=1200\,$MeV (2SC at $T=0$, left) and $\mu_B=1500\,$MeV (CFL at $T=0$, right). In the 2SC phase, $\Delta_1=\Delta_2=0$.}
    \label{fig:delta_t}
\end{figure*}

\noindent
In \cref{fig:delta_t} we compare the parameterized temperature dependence of the diquark gap using \cref{eq:Delta_T_approx} with the $ud$ pairing gap $\Delta_3(T=0)$ as input with the temperature dependence of the gaps calculated in the NJL model with $G_D=G_S$ for the 2SC phase and the CFL phase. In both cases, the critical temperature at which $\Delta_3$ vanishes, corresponding to the melting of up-down quark pairing, estimated from \cref{eq:Tc_approx} is about 10\,MeV smaller than the NJL value. In the CFL phase, the parameterized form lies between $\Delta_1(T)=\Delta_2(T)$ and $\Delta_3(T)$, representing a simpler, average description.

\section{Extrapolation of the pressure of melting quasiparticles}\label{app:extrapolation}

\begin{figure*}
      \includegraphics[width=0.5\linewidth]{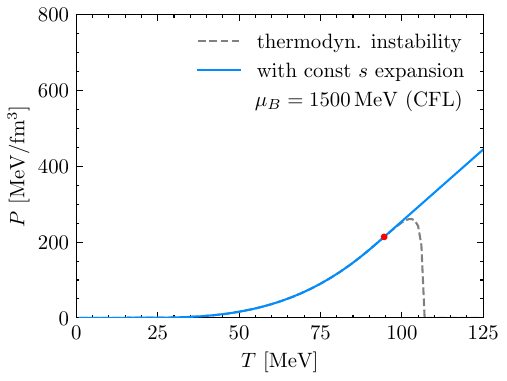}
     \caption{The pressure $P^\Delta$ (\cref{eq:quasiparticle_naive}) of quasiparticles with the temperature parameterization \cref{eq:Delta_T_approx} of the diquark condensate (gray, dashed) and the expression with a linear extrapolation \cref{eq:extrapolation_app} (blue solid) beyond the unstable point $T^\ast$ (red dot) plotted over temperature for the CFL phase at $\mu_B=1500\,$MeV. 
      }
     \label{fig:instability}
\end{figure*}

With the temperature dependence $\Delta(T)$ from \cref{eq:Delta_T_approx} we can write
\begin{equation}
    \frac{\Delta}{T}\equiv g(\tau)=\theta(\tau<1)\frac{\alpha}{\tau} ((1-\tau)^{3.4})^{0.53}
\end{equation}
with $\tau\equiv T/T_c$
which we plug into \eqref{eq:paired_app} to obtain
\begin{align}\label{eq:quasiparticle_naive}
    P^\Delta=
    \frac{\bar{\mu}^2}{\pi^{3/2}}T^2 f(\tau)
\end{align}
with the scaling function $f=h\circ g$ and 
\begin{align}
    h(x)\coloneqq g_{\text{2SC}}\sqrt{x} \left(\sqrt{2}-\frac{1}{2}e^{-x}\right).
\end{align}
for the 2SC phase and an analogous sum of the octet and singlet terms for the CFL phase, see \cref{sec:paired_expansion} and \cref{tab:T_exp}.
Note that the quasiparticle pressure in the form \cref{eq:quasiparticle_naive} melts to zero for $\tau>1$, corresponding to $T>T_c$. There is a temperature $T^\ast<T_c $, at which the second temperature derivative $\partial^2P^\text{paired}_{\text{th}}/\partial T^2$ becomes negative, indicating a decreasing entropy density with increasing temperature resulting in thermodynamic unstable matter.

In reality, however, the melting of the diquarks leads to unpaired quarks which provide thermal pressure that is not included in \cref{eq:quasiparticle_naive}. Our approach is targeted to work at low temperatures and we aim not to accurately describe the effect of diquark melting around the critical temperature. However, to avoid thermodynamic instability at $T^\ast$ we extrapolate \cref{eq:quasiparticle_naive} beyond $T^\ast$ using a constant entropy density approximation as in \cref{eq:extrapolation_generic}.\\
The temperature at which the instability happens is at
\begin{align}
    \frac{\partial^2  P^\Delta}{\partial T^2}\Bigg\vert_{T=T^\ast} =  \frac{\bar{\mu}^2}{\pi^{3/2}} \left(\tau^2 f''(\tau)+ 4\tau f'(\tau) + 2f(\tau)\right)\vert_{\tau=\tau^\ast} =0
\end{align}
which we find numerically using \texttt{MATHEMATICA}, to be at
\begin{equation}
    \tau^\ast \simeq \begin{cases}
        0.9039 \quad (\text{2SC}) \\
        0.8889 \quad (\text{CFL}) 
    \end{cases}
\end{equation}
for the 2SC and the CFL phase, respectively, i.e. the stability happens only shortly before the phase transition. For the entropy density at the point of the instability $\tau^\ast$ we find
\begin{equation}
    s(\tau^\ast,\mu_B) = \frac{\partial P^\text{paired}_{\text{th}}}{\partial T}\Bigg\vert_{T=T^\ast}
    = \frac{\bar{\mu}^2}{\pi^{3/2}}T^\ast\left(\tau^\ast f'(\tau^\ast)+2f(\tau^\ast)\right)
    \simeq \begin{cases}
        8.944  \dfrac{\bar{\mu}^2}{\pi^{3/2}}T^\ast \quad (\text{2SC}) \\\\
        12.09 \dfrac{\bar{\mu}^2}{\pi^{3/2}}T^\ast\quad (\text{CFL}) 
    \end{cases}.
\end{equation}
This is the constant slope which we will apply to model the thermal pressure of quasiparticles after the point of the instability $\tau^\ast$:
\begin{equation}\label{eq:extrapolation_app}
    P^{\text{quasip.}}(T,\mu_B)=\begin{cases}
        P^\text{paired}_{\text{th}}(T,\mu_B)\,, \quad T< T^\ast\\
         P^\text{paired}_{\text{th}}(T^\ast,\mu_B) + s(\tau^\ast,\mu_B) (T-T^\ast)\,, \quad T> T^\ast.
    \end{cases}
\end{equation}
The result of this procedure is shown for the CFL phase at $\mu_B=1500\,$MeV in \cref{fig:instability} where we compare the thermal pressure of the quasiparticles using \cref{eq:quasiparticle_naive} with the result \cref{eq:extrapolation_app} that uses a constant entropy-density extrapolation beyond $T^\ast$. Without the correction, the thermal pressure becomes unstable at $T^\ast$, indicated by the red dot in \cref{fig:instability}, and goes to zero at higher temperatures. The constant-entropy density extrapolation \cref{eq:extrapolation_app} mitigates this problem and is easy to implement.

\section{Thermodynamic stability at finite Temperature}\label{app:generic_stability}

Here we discuss the case that negative higher order coefficients appear in the Taylor expansion, confer the plots for $c_2$ and $c_3$ in \cref{fig:dsdt23_T=0}. We assume that the Taylor expansion is done at a fixed vector $\vec{\mu}$ of chemical potentials, as in Ref.~\cite{Mroczek:2024sfp}. In this case, we have to check that $\frac{\partial s}{\partial T}\Big|_{\vec{\mu}}(T,\vec{\mu})$ remains positive, which would otherwise imply a negative specific heat. The entropy density at finite $T$ is: 
\begin{equation}
s(T,\vec{\mu})=\frac{\partial s(T,\vec{\mu})}{\partial T}\biggr\rvert_{T=0,\vec{\mu}} T+\frac{1}{2}\frac{\partial s^2(T,\vec{\mu})}{\partial T^2}\biggr\rvert_{T=0,\vec{\mu}} T^2+\frac{1}{6}\frac{\partial^3 s(T,\vec{\mu})}{\partial T^3}\biggr\rvert_{T=0,\vec{\mu}} T^3+\mathcal{O}(T^4).
\end{equation}
We can then determine the heat capacity at any finite $T$ up to $\mathcal{O}(T^2)$ by taking a temperature derivative:
\begin{equation}
    \frac{\partial s}{\partial T}\Big|_{\vec{\mu}}(T,\vec{\mu})=\frac{\partial s(T,\vec{\mu})}{\partial T}\biggr\rvert_{T=0,\vec{\mu}} +\frac{\partial s^2(T,\vec{\mu})}{\partial T^2}\biggr\rvert_{T=0,\vec{\mu}} T+\frac{1}{2}\frac{\partial^3 s(T,\vec{\mu})}{\partial T^3}\biggr\rvert_{T=0,\vec{\mu}} T^2+\mathcal{O}(T^3)
\end{equation}
or written in terms of our $c_n$ coefficients for compactness:
\begin{equation}
    \frac{\partial s}{\partial T}\Big|_{\vec{\mu}}(T,\vec{\mu})=c_2 +c_3 T+\frac{1}{2}c_4 T^2+\mathcal{O}(T^3).
\end{equation}
For thermodynamic stability we require that $ \frac{\partial s}{\partial T}|_{\vec{\mu}}(T,\vec{\mu})\geq 0$ for all $T$ and $\vec{\mu}$, which also places a constraint that $c_2\geq 0$. Unlike $c_2$, we have already shown in this work that $c_3$ and $c_4$ may be either positive or negative such that it is conceivable that at a certain $T$ the positivity of $ \frac{\partial s}{\partial T}|_{\vec{\mu}}(T,\vec{\mu})$ is no longer guaranteed if only terms up to $c_4$ are kept in the expansion.

We can find the critical temperature $T^*$ where $ \frac{\partial s}{\partial T}|_{\vec{\mu}}(T,\vec{\mu})$ switches sign by solving:
\begin{equation}
   0=c_2 +c_3 \cdot T^*+\frac{1}{2}c_4 \cdot (T^*)^2
\end{equation}
from which we obtain
\begin{equation}
    T^*=\frac{-c_3\pm \sqrt{c_3^2-2c_2c_4}}{c_4}.
\end{equation}

We point out certain interesting limits of this equation. For instance, in some phases of matter there is a hierarchy of coefficients such that $c_2\gg c_3\gg c_4$. In the case of the hierarchy, we could drop the $c_4$ term entirely and
\begin{equation}
    T^*=-\frac{c_2}{c_3}
\end{equation}
where we find that we only have to worry about this temperature if $c_3<0 $, otherwise the sign change would only occur for $T^*<0$ which we never encounter. Furthermore, since $c_2\gg c_3$ we expect $T^*$ to be quite large. For example, in both this work and \cite{Mroczek:2024sfp} there's often at least 2 orders of magnitude difference between $c_2$ and $c_3$ such that $T^*\gtrsim 100$ MeV in this limit.

Next, let's study the limit of conformal EoS that have vanishing $c_3\sim 0$ but large $c_4$ terms. We then obtain:
\begin{equation}
    T^*=\pm\sqrt{-2\frac{c_2}{c_4}}.
\end{equation}
In this limit, we only have to worry about this sign change if $c_4<0$.
However, if $c_4$ is approximately two orders of magnitude smaller than $c_2 T^2$ than we already arrive at issues at much lower temperatures. For instance, if $c_2 (T^\ast)^2/c_4\sim -100$ we find that $T^*\sim 14$ MeV, which is quite low. 
Thus, a major challenge to our finite $T$ expansion is large, negative $c_4$ terms. 
These terms are somewhat unlikely but may occur around phase transitions or crossovers (as we see in Fig.\ \ref{fig:dsdt23_T=0}). We also note that higher or lower order $c_n$ terms may cancel this effect, so this remains to be studied further. 

Regardless of the exact value of $T^*$, we propose the following method to handle the sign change of $ \frac{\partial s}{\partial T}|_{\vec{\mu}}(T,\vec{\mu})$. For $T\geq T^\ast$ one can linearly extrapolate $P(T,\vec{\mu})$ with the slope given by then entropy density at the point of the instability $s(T^\ast,\vec{\mu})$:
\begin{equation}\label{eq:extrapolation_generic}
    P(T,\vec{\mu})=\begin{cases}
        P^{\text{Taylor}}(T,\vec{\mu}), \quad T< T^\ast\\
         P(T^\ast,\vec{\mu}) +  (T-T^\ast)s(T^\ast,\vec{\mu}) , \quad T> T^\ast.
    \end{cases}
\end{equation}
This result now provides a thermodynamically stable pressure at temperatures higher than $T^*$.

\section{Derivatives of the entropy density in the NJL model}\label{app:derivatives}
The full expression for the pressure of the NJL model in the RG-consistent mean-field approximation is given by \cite{Gholami:2024diy}
\begin{align}\label{eq:pressure_full}
P(T,\mu_B)
=& -\mathcal{V}(\bar{\bm{\chi}}) 
+ \frac{1}{2\pi^2} \int_0^{\Lambda} dp\, p^2 \mathcal{A}(T,\mu_B,\bar{\mu}_3,\bar{\mu}_8,\bar{\bm{\chi}}) \\
&- \frac{1}{2\pi^2}\int_{\Lambda'}^{\Lambda} dp\, p^2 \Bigg(\mathcal{A}(T=0,\mu_B=0,\mu_3=0,\mu_8=0,\bar{\bm{\chi}}) \\
&+  \sum \frac{1}{2}\mu_{\alpha a,\beta b}^2 \cdot
\left(\frac{\partial^2}{\partial \mu_{\alpha a,\beta b}^2} \mathcal{A}(T=0,\mu_B,\mu_3,\mu_8,\bar{\bm{\chi}})\right) \bigg|_{\mu_B=\mu_3=\mu_8=0;\Delta_{\alpha a,\beta b}\neq 0}\Bigg)
\end{align}
with the momentum integrand
\begin{equation}
\mathcal{A}(T,\mu_B,\bar{\mu}_3,\bar{\mu}_8,\bar{\bm{\chi}})=\sum_{j=1}^{18}\left(
    \epsilon_j(\bm{p};\mu_B,\bar{\mu}_3,\bar{\mu}_8,\bar{\bm{\chi}}) + 2T\ln\left(
   1+e^{-\frac{\epsilon_j(\bm{p},\mu_B,\bar{\mu}_3,\bar{\mu}_8,\bar{\bm{\chi}})}{T}}\right)\right).
\end{equation}
Here, $\bar{\bm{\chi}}=\{\bar{\phi}_u,\bar{\phi}_d,\bar{\phi}_s,\bar{\Delta}_1,\bar{\Delta}_2,\bar{\Delta}_3\}$ denotes the chiral condensates and diquark condensates, respectively, set to their physical mean-field value, $\bar{\mu}_3$, $\bar{\mu}_8$ are the values of the color chemical potentials set by charge neutrality and $\epsilon_j$ are positive eigenvalues of the inverse quark propagator, for more details see Ref.~\cite{Gholami:2024diy}.

For the calculation of naive derivatives for the finite $T$ expansion, only the term with explicit temperature dependence, i.e. the second term in the first line of \cref{eq:pressure_full} is relevant. The entropy density $s$ and its naive derivatives are given by
\begin{alignat}{2}
   c_1=& s &&= \frac{1}{\pi^2} \int_0^\Lambda \sum_{j=1}^{18} \left(\ln(1+\exp(-\epsilon_j/T) + \frac{\epsilon_j}{T \cdot (1+\exp(\epsilon_j/T))}\right)\,,\\
      c_2=& \frac{\partial s}{\partial T}\Big\vert_{T,\mu_B}^{\text{naive}}&&=
       \frac{1}{\pi^2} \int_0^{\Lambda} dp\, p^2 \sum_{j=1}^{18}  \frac{\epsilon_j^2}{T^3} \frac{\exp({-\epsilon_j/T})}{(1+\exp{(-\epsilon_j/T)})^2}\,,\\
      c_3=& \frac{\partial^2 s}{\partial T^2}\Big\vert_{T,\mu_B}^{\text{naive}} &&=  \frac{1}{\pi^2}  \int_0^{\Lambda} dp\, p^2 \sum_{j=1}^{18} \frac{\epsilon_j^2 \cdot \text{sech}^2(\epsilon_j/(2T))}{4T^5}(-3T+\epsilon_j\cdot\tanh{(\epsilon_j/(2T))})\,,\\
       c_4=&\frac{\partial^3 s}{\partial T^3}\Big\vert_{T,\mu_B}^{\text{naive}} &&=   \frac{1}{\pi^2}  \int_0^{\Lambda} dp\, p^2 \sum_{j=1}^{18} \frac{1}{8T^7} \left( 
       \epsilon_j^2 \text{sech}^4(\epsilon_j/(2T))\cdot(12T^2 -2 \epsilon_j^2+(12T^2+\epsilon_j^2)\cdot \cosh(\epsilon_j/T)-8T\epsilon_j\sinh(\epsilon_j/T))
       \right).
\end{alignat}
The first two coefficients $c_1$ and $c_2$ are always non-negative, however, the integrand of $c_3$ changes sign at $\epsilon_j\approx \pi T$, see the expression in parentheses. The resulting value is therefore much smaller in absolute value than $c_2\cdot T$ for the temperatures of interest and can be positive or negative, see \cref{fig:dsdt23_T=0}.

\section{Handling of 2SC-CFL transition at nonzero temperature}\label{app:smooth}

\begin{figure}[h]
      \includegraphics[width=0.5\linewidth]{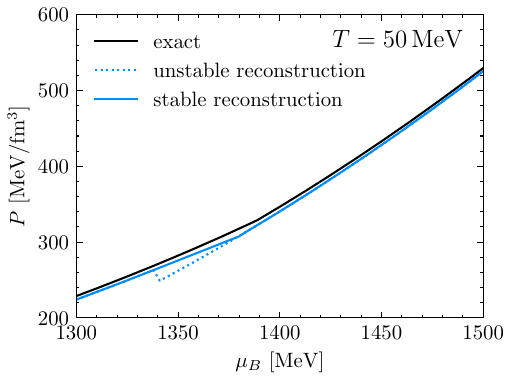}
     \caption{Pressure versus baryon chemical potential at $T=50$ in the model with diquark pairing ($G_D=G_S$). We compare the exact mean-field result in black, with the reconstruction given by \cref{eq:general_pairings}. Applying the construction naively leads to unphysical instabilities in the EoS after the 2SC-CFL phase transition (blue, dotted) which can be solved via interpolation (blue, solid).
      }
     \label{fig:smooth}
\end{figure}

In the phase diagram of our model with diquark pairing (\cref{fig:composition}, right), the 2SC-CFL phase boundary is located at $\mu_B\approx1340\,$MeV at zero temperature and bends towards higher $\mu_B$ with increasing temperature. At higher chemical potential $\mu_B\gtrsim1340\,$MeV there are points in the phase diagram where matter is in the 2SC phase at nonzero temperature but in the CFL phase at zero temperature. Our reconstructed EoS only uses $T=0$ information, thus for these points the CFL expansion formula is used. This can create unphysical non-monotonic behavior in the pressure as a function of the chemical potential at a fixed temperature. More specifically, a dip in the pressure can appear, at which the number density $n_B=\partial P /\partial \mu_B\vert_T$ is negative, see the dotted blue line for $T=50\,$MeV in \cref{fig:smooth}.\\
To remove this artifact which is due to insufficient knowledge about the phase boundary at finite temperature from $T=0$ information, we linearily extrapolate the pressure of the 2SC branch right at the start of the unphysical dip until its intersection to the CFL branch, which makes the pressure a monotonically increasing function of $\mu$, see the solid light blue curve in \cref{fig:smooth}. This way, we obtain a  kink in the pressure (around $\mu_B\simeq 1380\,$MeV) characteristic of a first order phase transition while $P(\mu_B)$ monotonically increases (at a fixed $T$). Note that this approach leads to an effective first-order 2SC-CFL transition even at higher temperatures, where the phase transition in the true model is second order.\\

\twocolumngrid

\bibliography{inspire.bib,NOTinspire.bib}
\bibliographystyle{apsrev4-2}

\end{document}